\documentclass[%
reprint,
superscriptaddress,
longbibliography,
 amsmath,amssymb,
 aps,
prb,
floatfix,
]{revtex4-2}
\usepackage[english]{babel}
\usepackage{comment}
\usepackage{amsmath}
\usepackage{natbib}
\usepackage{ulem}
\usepackage{graphicx}
\usepackage{dcolumn}
\usepackage{bm}
\usepackage{amsmath}
\usepackage{siunitx}
\usepackage{listings}
\usepackage{mathtools}
\usepackage{float}
\usepackage{dsfont}

\usepackage{xcolor}
\usepackage{url}
\usepackage{hyperref}
\hypersetup{
    colorlinks=true,allcolors=blue,}

\begin{document}
\preprint{APS/123-QED}

\title{Majorana bound states in a $d$-wave superconductor planar Josephson junction}
\author{Hamed Vakili}
\affiliation{Department of Physics and Astronomy and Nebraska Center for Materials and Nanoscience, University of Nebraska, Lincoln, Nebraska 68588, USA}
\author{Moaz Ali}
\affiliation{Department of Physics and Astronomy and Nebraska Center for Materials and Nanoscience, University of Nebraska, Lincoln, Nebraska 68588, USA}
\author{Mohamed Elekhtiar}
\affiliation{Department of Physics and Astronomy and Nebraska Center for Materials and Nanoscience, University of Nebraska, Lincoln, Nebraska 68588, USA}
\author{Alexey A. Kovalev}
\affiliation{Department of Physics and Astronomy and Nebraska Center for Materials and Nanoscience, University of Nebraska, Lincoln, Nebraska 68588, USA}
\date{\today}

\begin{abstract}
We study phase-controlled planar Josephson junctions comprising a two-dimensional electron gas with strong spin-orbit coupling and $d$-wave superconductors, which have an advantage of high critical temperature. We show that a region between the two superconductors can be tuned into a topological state by the in-plane Zeeman field, and can host Majorana bound states. The phase diagram as a function of the Zeeman field, chemical potential, and the phase difference between superconductors exhibits the appearance of Majorana bound states for a wide range of parameters. We further investigate the behavior of the topological gap and its dependence on the type of $d$-wave pairing, i.e., $d$, $d+is$, or $d+id^\prime$, and note the difficulties that can arise due to the presence of gapless excitations in pure $d$-wave superconductors. On the other hand, the planar Josephson junctions based on superconductors with $d+is$ and $d+id^\prime$ pairings can potentially lead to realizations of Majorana bound states. Our proposal can be realized in cuprate superconductors, e.g., in a twisted bilayer, combined with the layered semiconductor Bi$_2$O$_2$Se. 
\end{abstract}
\pacs{Valid PACS appear here}
\maketitle 

Majorana bound states (MBS) are non-Abelian anyons that can be used in realizations of quantum computers relying on topological protection~\cite{Nayak2008,Kitaev2003}. Anyons are quasiparticles that are described by the statistics of neither fermions nor bosons and have exotic properties such as fractional charge~\cite{Nayak2008,Kitaev2003}. Topological protection associated with realizations of non-Abelian anyons in condensed matter systems can be used to encode quantum information in a way that is robust against decoherence~\cite{Alicea2012,Beenakker2013,Elliott2015}. The current common platforms to realize MBS are relying on proximity effect~\cite{FuKane,Eschrig2019,proximitylevin,Zutic2019,Gngrd2022} with $s$-wave superconductors which have low critical temperatures~\cite{Alicea2012,Beenakker2013,Elliott2015}. The topological gap in such realizations is relatively small, making the system more sensitive to disorder~\cite{Takei2013,Zhang2019} and various imperfections~\cite{Rainis2012,Stanescu2014}, and requiring operation at very low temperatures~\cite{pikulin_protocol_2021,aghaee_inas-hybrid_2022,Huo_2023}.

The $d$-wave superconductivity is a very common type of superconductivity in strongly correlated systems such as cuprates~\cite{Yanase2003}. Such superconductors are associated with gapless excitations~\cite{Sato2017} and higher critical temperatures compared to $s$-wave superconductors. Unfortunately, the presence of gapless excitations is not compatible with topological superconductivity. Furthermore, in $d$-wave superconductors, gapless excitations lead to the appearance of Andreev bound states (ABS)~\cite{Adagideli1999,Chen2001,Sengupta2001,Tanaka2012}. 
To resolve issues associated with gapless excitations, one can use $d$-wave superconductors with inversion asymmetry and an applied magnetic field~\cite{MZMTI,dwave2deg,Liu2018,magnetchain,Yanase2022}, or use a twisted bilayer of $d$-wave superconductors that behaves as an effective $d+id'$ superconductor~\cite{Liu2018,Crawford2020,Mercado2022,Margalit2022,Can2021}. As has been demonstrated in Ref.~\cite{Can2021}, a twisted bilayer of $d$-wave superconductors offers a high degree of tunability via variations in twist angle and also offers the realization of $d+is$ or $d+id^\prime$ pairings within the same system. The planar Josephson junction (JJ) platform based on $s$-wave superconductors allows to realize robust and tunable MBS~\cite{JJ2017,Halperin2017,Fornieri2019,narrowJJ,Zhou2022}. Recently, there has been substantial interest in various realizations of a planar JJ based on $s$-wave superconductors in the context of a superconducting diode effect~\cite{Ando2020,Mayer2020,Dartiailh2021,Baumgartner2021,Amundsen2022}. It has been demonstrated that the superconducting diode effect also appears in a JJ based on $d$-wave superconductors and a topological insulator~\cite{Tanaka2022}. 

\begin{figure}[h]
    \centering
    \includegraphics[width=\columnwidth]{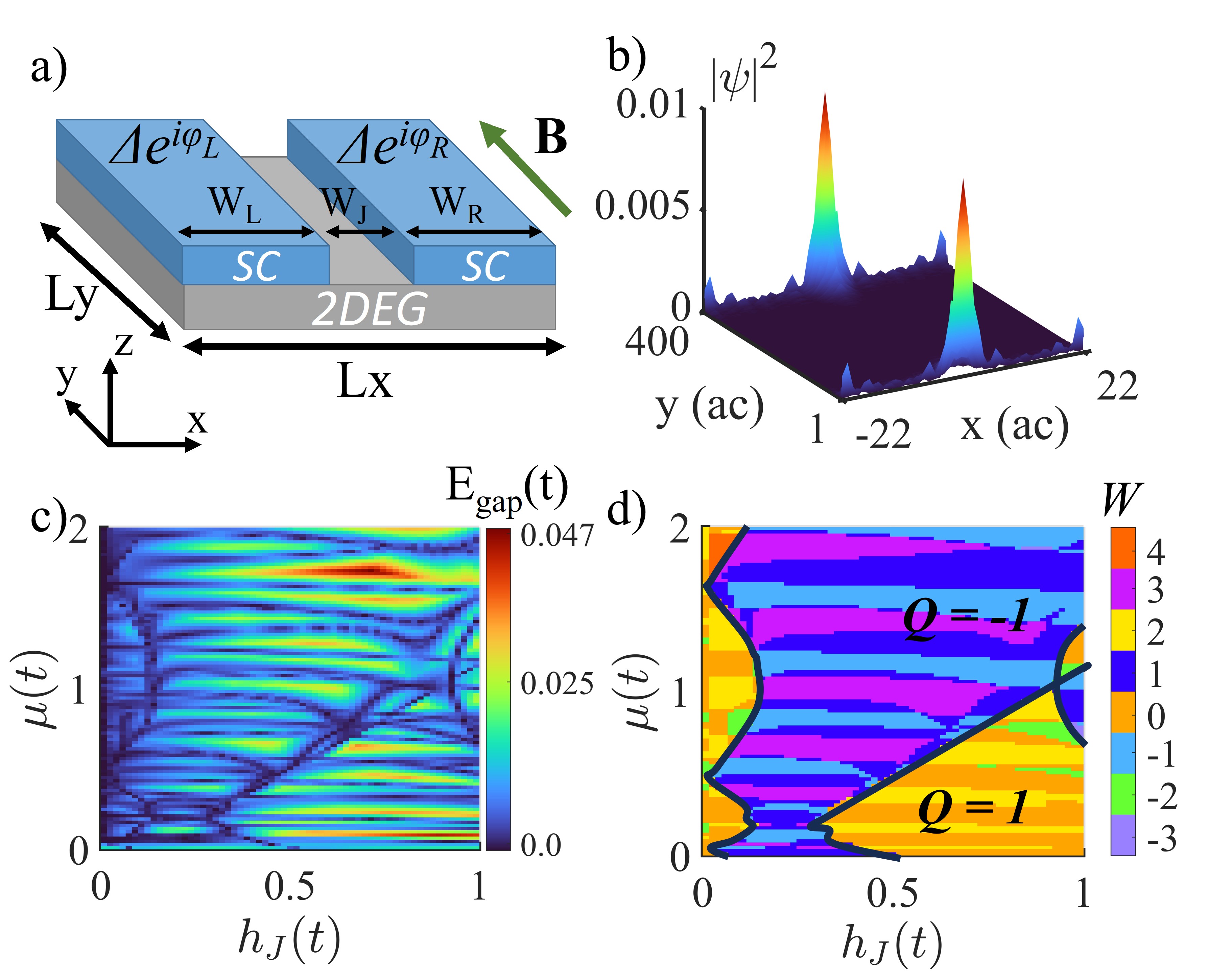}
    \caption{a) Schematic diagram of the planar Josephson junction heterostructure. The 2DEG is covered by $d$-wave superconductors, except for a junction of width $W_J$ between the two superconductors. In b) the probability function ($|\psi|^2$) of MBS at $\mu = 1.32t$ and $h_J = 0.6t$ is plotted. c) The energy gap as a function of the Zeeman field $h_J$ in the junction and $\mu$. d) The topological invariants $Q$ ($\mathbb{Z}_2$) and $W$ ($\mathbb{Z}$) as a function of of the Zeeman field $h_J$ in the junction and $\mu$. The colors show the value of the $W$ invariant. The black lines separate the regions with $Q = $ -1 and 1. The $W$ invariant only changes between odd (even) values for $Q = $ -1 (1) regions.
    We used the parameters: $L_y = 400a_c$, $W_J = 5a_c$, $W_L = W_R = 20a_c$, $\Delta_0 = 0.3t$, $\Delta^\prime_0 = 0.0 $, $h_{SC} = 0$, and the phase difference $\Delta\phi$ = $\pi$.}
    \label{fig:MBS}
\end{figure}

In this Letter, we demonstrate that a planar JJ comprising a two-dimensional electron gas (2DEG) with strong spin-orbit coupling and $d$-wave superconductors can host MBS (see Fig.~\ref{fig:MBS}a)). Such systems can be realized using
exfoliated copper oxide heterostructures~\cite{Yu2019,Can2021} combined with a layered semiconductor, such as Bi$_2$O$_2$Se~\cite{Wu2017,Margalit2022}. The presence of the proximity effect can result in the pairing terms in the neighboring semiconductor~\cite{Margalit2022}. In the case of pure $d$-wave pairing in 2DEG, we observe that MBS coexist with gappless bulk states which can be detrimental for storing quantum information. However, for the $d+is$ and $d+id^\prime$ pairings in 2DEG, there exists a bulk gap and MBS arise in the planar JJ geometry. The $d+is$ pairing leads to realizations of robust MBS in analogy to JJ based on $s$-wave superconductors~\cite{Halperin2017}. For $d+id^\prime$ pairing, in addition to MBS, the chiral edge modes exist at zero energy in the proximitized 2DEG due to nontrivial bulk topology; however, they can be gapped out inside JJ by properly tuning the Zeeman field, the relative phase of superconductors, and the chemical potential. As a result for $d+id^\prime$ pairing, one can physically separate the MBS and the chiral edge modes, as we show in our study.
 

\textit{Model and symmetry analysis.~} We consider the planar JJ geometry with an in-plane Zeeman field (due to a magnetic field or induced by proximity with a ferromagnet), and mostly assume the limit of large superconductors. When necessary, we use periodic boundaries along the 
$y$ axis in Fig.~\ref{fig:MBS}a). The phase difference of the left and right $d$-wave superconductors can be controlled by applying current or magnetic flux.
The Bogoliubov–de Gennes (BdG) Hamiltonian of the system, written in the Nambu basis ($\psi_\uparrow,\psi_\downarrow,\psi^\dagger_\downarrow,-\psi^\dagger_\uparrow$), reads 
\begin{align}
H &=\left(\frac{\boldsymbol{p}^2}{2m^*} -\mu+\frac{\alpha_{SO}}{\hbar}\left(\boldsymbol{\sigma} \times \boldsymbol{p}\right)_z +\frac{m^*}{2}\alpha_{SO}^2\right) \tau_{z}\nonumber\\  
&+ h(x) \sigma_y+\tau_x \text{Re}\Delta(\boldsymbol k,x) -\tau_y\text{Im}\Delta(\boldsymbol k,x).\label{eq:Hamiltonian}
\end{align}
Here, $\alpha$ is the Rashba spin-orbit coupling strength, $h(x)$ describes the Zeeman energy, e.g., for an external magnetic field $h(x)=g(x)\mu_B B/2$, where $g(x)$ is the position-dependent $g$-factor. We denote the Zeeman energy in the superconducting regions by $h_{SC}$ and in the junction by $h_J$, $m^*$ is the effective electron mass, and $\Delta(\boldsymbol k,x)$ is the proximity-induced pairing in 2DEG, which reads
\begin{align}
    \Delta(\boldsymbol k,x) &=  2e^{i\phi(x)} [\Tilde{\Delta}(x)\left(\cos{k_y}- e^{i\beta}\cos{k_x}\right)\nonumber \\ &+\Tilde{\Delta}^\prime(x) 2i \sin{k_x} \sin{k_y}].
\end{align}
Here, $\beta$ determines the pairing type, e.g., $\beta = 0$ corresponds to $d_{x^2-y^2}$ pairing and $\beta =\pi/4$  corresponds to $d+is$ pairing~\cite{Can2021}. The term $\Tilde{\Delta}^\prime$ is necessary to realize $d+id^\prime$ pairing. The terms $\Tilde{\Delta}(x)$ and $\Tilde{\Delta}^\prime(x)$ are only nonzero in the regions covered by superconductors where $\Tilde{\Delta}(x)=\Delta_0$, $\Tilde{\Delta}^\prime(x)=\Delta_0^\prime$, $\phi(x)=\phi_L$ for the region covered by the left superconductor, and $\phi(x)=\phi_R$ for the region covered by the right superconductor. We apply the Zeeman field in the $y$ direction. 

The symmetries, such as the time-reversal symmetry, the particle-hole symmetry, and the chiral symmetry determine the type of topological superconductivity realizable in our system.
The Hamiltonian~\eqref{eq:Hamiltonian} breaks time-reversal symmetry when the Zeeman field is present. For the case of $d$ pairing (or $d+id'$ pairing when $\Delta^\prime \neq 0$) in the presence of the Zeeman field, we can still define an effective time-reversal symmetry as $\Tilde{T}=M_x T$ where $T=i\sigma_y K$ is the usual time-reversal symmetry and $M_x$ is the mirror with respect to the $y\mbox{-}z$ plane. For $\Tilde{T}$, we have a relation $\Tilde{T}^2=1$, which in combination with the particle-hole symmetry, $P=\sigma_y\tau_y K$, and the chiral symmetry (the $P\Tilde{T}$ symmetry operator), $\Tilde{C}=M_x \tau_y$, places such a system in the BDI symmetry class with a $\mathbb{Z}$ topological invariant. In our analysis, we also calculate a $\mathbb{Z}_2$ Pfaffian invariant which is determined by the parity of $\mathbb{Z}$ and characterizes systems in the D symmetry class. The D symmetry class is realized in Fig.~\ref{fig:MBS} when the mirror symmetry with respect to the $y\mbox{-}z$ plane is broken or for $d+is$ pairing.
 
\textit{Topological superconductivity and MBS.} ~Our symmetry analysis suggests a possibility for topological superconductivity in the quasi-one-dimensional region between superconductors in Fig.~\ref{fig:MBS}a). A sufficiently large but finite system along the $y$ axis can be used for realizing MBS. We use the tight-binding version of the BdG Hamiltonian \eqref{eq:Hamiltonian} on a two-dimensional square lattice given by:
\begin{align}
    H_{TB} = &\sum_{<ij,mn>}\left[-t ~c^\dagger_{i,j} c_{m,n} + it_{SO}~c^\dagger_{i,j}\left( \boldsymbol{\sigma} \times \mathbf{r}_{ij,mn}\right)_z c_{m,n}\right]\nonumber\\
    &+ \sum_{i,j}\left[(4t-\mu+t_{SO}^2) ~c^\dagger_{i,j} c_{i,j}+ h~c^\dagger_{i,j} \sigma_y~c_{i,j}   \right]\nonumber\\
    &+ \sum_{i,j}e^{i\phi/2}\Tilde{\Delta}\left[~c^\dagger_{i,j\pm 1,\uparrow} c^\dagger_{i,j,\downarrow}-~e^{i\beta}c^\dagger_{i\pm 1,j,\uparrow} c^\dagger_{i,j,\downarrow}\right]+\nonumber\\
    &\sum_{i,j}e^{i\phi/2}\Tilde{\Delta}^\prime\left[~c^\dagger_{i\pm1,j\pm1,\uparrow} c^\dagger_{i,j,\downarrow}-~c^\dagger_{i\pm1,j\mp 1,\uparrow} c^\dagger_{i,j,\downarrow}\right]\nonumber\\
    & + H.c., \label{eq:TBHamiltonian}
\end{align}
where in our calculations we use the lattice spacing $a_c = 10$~nm, $t_{SO}=0.3t$, and $t = \frac{\hbar^2}{2m^* a_c^2}$, and $m^*$= $0.14 m_e$~\cite{Wu2017}, with $m_e$ being the electron rest mass. We study the phase diagram as a function of the Zeeman field, chemical potential, and the phases of superconductors. To describe a system with periodic boundaries along the $y$ axis, we use the momentum representation along the $y$-axis of the tight-binding Hamiltonian, $H_\text{TB}(k_y)$.
To calculate the BDI class  $\mathbb{Z}$ topological invariant, we employ the eigen-basis of the chiral symmetry in which $\Tilde{C}$ is diagonal~\cite{Sau2012,Halperin2017} with $\mathds{1}$ in the upper left block and $-\mathds{1}$ in the lower right block. In this basis,
\begin{equation}
\Tilde{H}_\text{TB}(k_y)=
\begin{pmatrix}
0 & A(k_y)\\
A^\dagger(k_y) & 0
\end{pmatrix},  
\end{equation}
where for a gapped Hamiltonian $A(k_y)$ defines a complex function $z(k_y)=Det(A(k_y))/|Det(A(k_y))|$ and a winding number $W=(-i/2\pi)\int_{k_y=0}^{k_y=2\pi}dz(k_y)/z(k_y)$. To calculate the D class $\mathbb{Z}_2$ topological invariant, we use the expression $Q = \text{sign}(\text{Pf}(H_{k_y=\pi}\sigma_y\tau_y)/\text{Pf}(H_{k_y=0}\sigma_y\tau_y))$.
The two topological numbers are related by the equation $(-1)^W = Q$~\cite{Sau2012}.

\textit{Planar JJ with $d$-wave pairing.} ~We first study the planar JJ in Fig.~\ref{fig:MBS}a) with pure $d$-wave pairing by varying the chemical potential, the phase difference ($\Delta\phi = \phi_R - \phi_L$), and the Zeeman field. We use parameters: $W_J$ = 5$a_c$, $W_L=W_R$ = 20$a_c$, $\Delta_0 = 0.3$t, $h_{SC}=0$, $\beta=0$, and $\Delta^\prime_0 = 0.0$.
The nontrivial topological regions in parameter space, as determined by $Q$, are reminiscent of the planar JJ based on the $s$-wave superconductor, which suggests the same mechanism for the formation of MBS and the relevance of the Andreev reflection~\cite{Halperin2017}.  
In Fig.~\ref{fig:MBS}b), we show the energy gap of the system as a function of $\mu$ and $h_J$. The gap shows rapid changes, exhibiting numerous lines with zero gap, which suggests that this is a finite size effect associated with the gapless excitations in the bulk of 2DEG. Further simulations with increased system sizes show an increase in the number of lines with zero gap. According to Fig.~\ref{fig:MBS}c), the gap closes along the lines that seem to be associated with changes in the two topological invariants, $W$ or $Q$, which are calculated from $H_\text{TB}(k_y)$. From a decrease of the gap for larger system sizes, we conclude that in the region with $Q = -1$ MBS will coexist with gapless excitations in the planar JJ based on the $d$-wave pairing. Even though the topological invariant $Q$ shows a large continuous region with $Q=-1$ and a well-localized MBS can be realized as shown in Fig.~\ref{fig:MBS}b), MBS may still hybridize with the gapless excitations. This behavior can be detrimental for quantum coherence associated with MBS. Nevertheless, we expect that the signature of the $Q = -1$ region can be seen in studies of the Josephson current and the superconducting diode effect.  

\begin{figure}[h]
    \centering
    \includegraphics[width=\columnwidth]{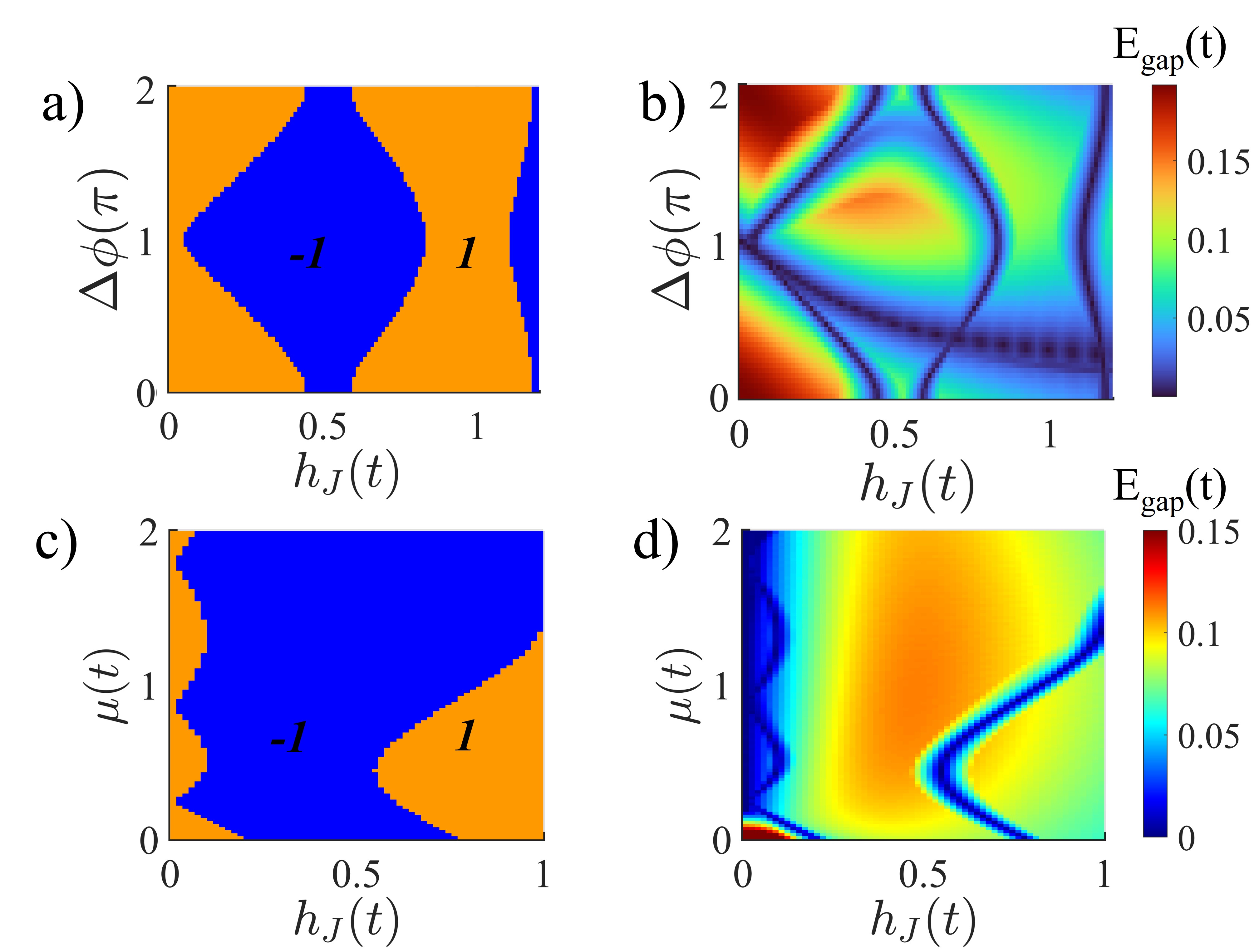}
    \caption{Phase diagrams for $d+is$ pairing ($\beta = \pi/4$). a) The $Q$ topological number ($\mathbb{Z}_2$) as a function of $h_J$ and $\Delta\phi$ for $\mu = 1t$.  b) The energy gap $E_{gap}$ as a function of $h_J$ and $\Delta\phi$ for $\mu = 1t$. [The dark blue line extending from left to right has a small gap, which is why no corresponding topological phase change exists in a)] c) Phase diagram of the $Q$ topological number as a function of $h_J$ and $\mu$ for the phase difference $\Delta\phi = \pi$. d) The energy gap $E_{gap}$ as a function of $h_J$ and $\mu$ for the phase difference $\Delta\phi = \pi$. We used the parameters: $h_{SC} = 0$, $\Delta_0 = 0.3 t$, $\Delta^\prime_0 = 0.0 $, $W_J$ = 5$a_c$, $W_L =  W_R = 40a_c$. }
    \label{fig:Z2}
\end{figure}
 
\begin{figure}
    \vspace{2mm}
    \centering
    \includegraphics[width=\columnwidth]{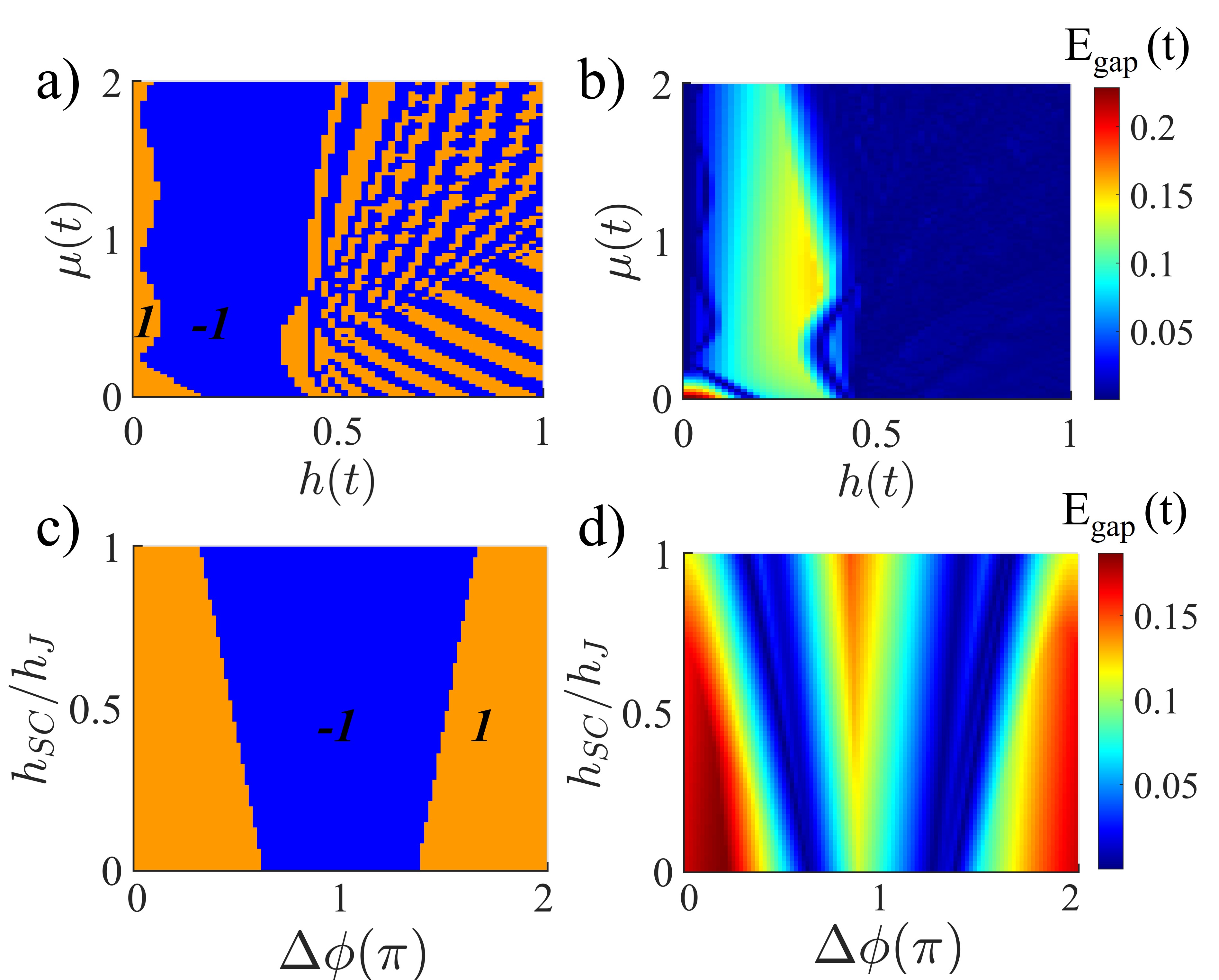}
    \caption{ The Zeeman field is applied to the whole system. a) The $Q$ topological invariant as a function of $h$ and $\mu$ where $h_J = h_{SC} = h$ for the phase difference $\Delta\phi = \pi$. The blue color shows the topological region ($Q=-1$). b) The energy gap $E_{gap}$ as a function of $h$ and $\mu$ where $h_J = h_{SC} = h$ for the phase difference $\Delta\phi = \pi$. c), d) The $Q$ topological number and the energy gap as a function of the ratio $h_{SC}/h_{J}$ and $\phi$ for $\mu = 1t$. We used the parameters: $\Delta_0 = 0.3 t$, $\Delta^\prime_0 = 0.0 $, $\beta = \pi/4$, $h_J=0.2t$, $W_J = 5a_c$, and $W_L = W_R = 40a_c$.  }
    \label{fig:coverB}
\end{figure}

\textit{Planar JJ with $d+is$ pairing.} ~Since $d$-wave superconductors have directional pairing resulting in $d_{x^2-y^2}$ and $d_{xy}$ components in a general reference frame, the orientation of the superconductor lattices can be used to control the pairing potential in twisted bilayer $d$-wave superconductors realizable in mechanically exfoliated copper oxide heterostructures. It has been shown~\cite{Can2021} that by having a twisted bilayer of $d$-wave superconductors, $d+is$, and $d+id^\prime$ pairings can be realized. In Fig.~\ref{fig:Z2}, we consider a system with $d+is$ pairing and use the parameters: $h_{SC} = 0$, $\Delta_0 = 0.3t$, $\Delta^\prime_0 = 0.0$, $W_J$ = 5$a_c$, $W_L=W_R$ = 40$a_c$. We take $\beta =\pi/4$ predicted by the BCS mean-field calculations~\cite{Can2021}. In Fig.~\ref{fig:Z2}a), we calculate the phase diagram for $Q$ as a function of $h_J$ and $\phi$. The large diamond-shaped region, also typical for the planar JJ based on the $s$-wave superconductor~\cite{Halperin2017}, defines parameters for which MBS can be realized. In Fig.~\ref{fig:Z2}b), we plot the gap. The gap closes at a line of quantum phase transition between the $Q=-1$ and $Q=1$ regions. In Figs.~\ref{fig:Z2}c) and d), we perform similar calculations as a function of $\mu$ and $h_J$ with fixed $\phi = \pi$. Results in Fig.~\ref{fig:Z2} predict robust MBS in the planar JJ with $d+is$ pairing for a wide range of parameters. We also observe in Fig.~\ref{fig:coverB} that the region with $Q=-1$ persists even when the Zeeman energy is present in the superconducting regions. Figure~\ref{fig:coverB}a) shows the phase diagram for $Q$ as a function of $\mu$ and $h_J = h_{SC} = h$. A relatively large region corresponding to $Q=-1$ in Fig.~\ref{fig:coverB}a) can lead to realizations of robust MBS due to a sizable topological gap shown in Fig.~\ref{fig:coverB}b). In Figs.~\ref{fig:coverB}c) and d), 
 we show that increasing the ratio $h_{SC}/h_{J}$ expands the range of $\phi$ where MBS can be realized for certain parameters. The imaginary $is$ pairing component is responsible for the gap opening here, which would also protect MBS from quasiparticle poisoning. Based on the BCS mean field calculations the $is$ component is $\tan(\pi/8)$ of the $d_{x^2-y^2}$ component. This would still be significantly larger than the gap in the conventional s-wave superconductors.

\begin{figure}
    \centering
    \includegraphics[width=\columnwidth]{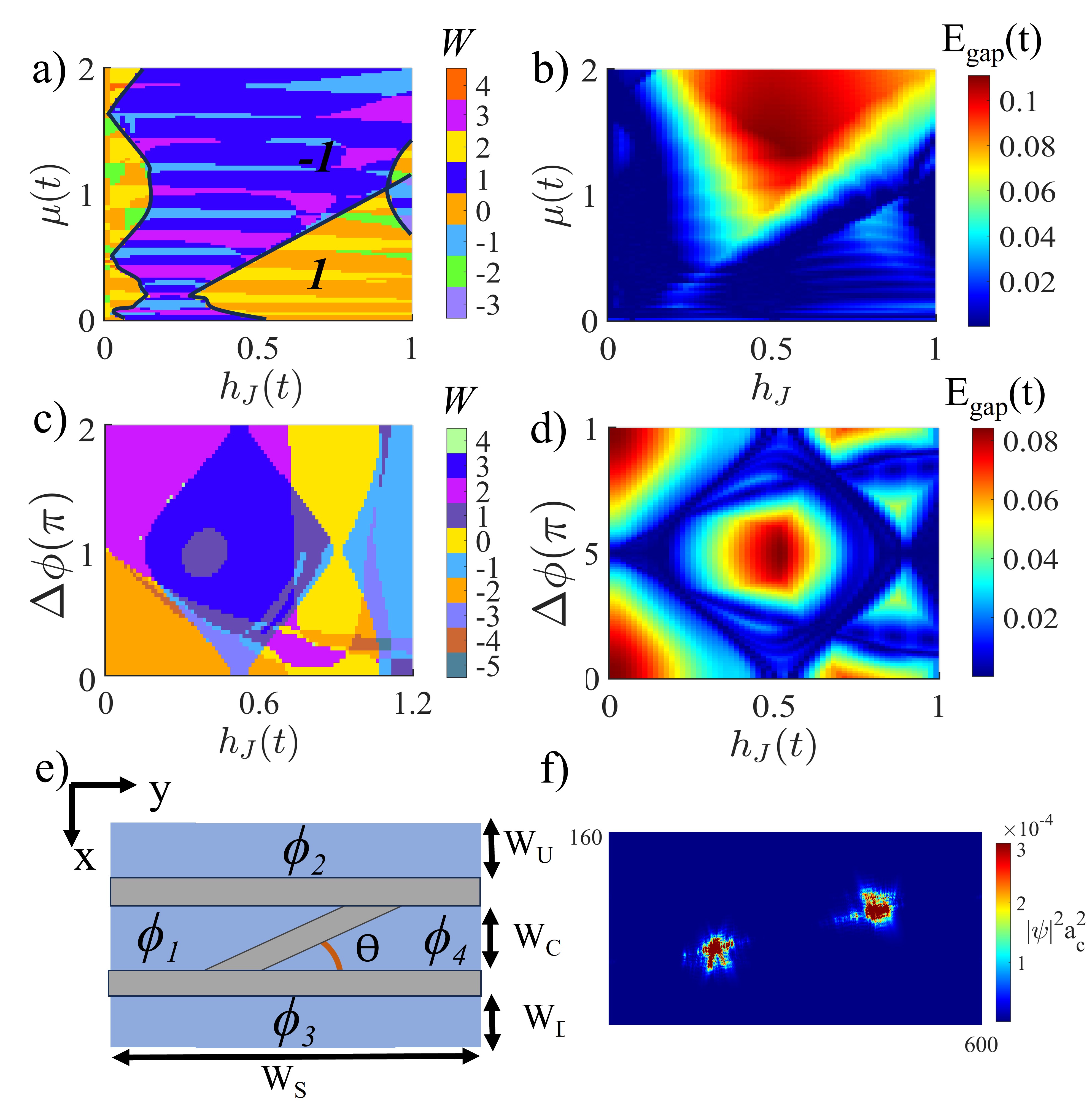}
    \caption{a) The topological invariant $W$ as a function of $h_J$ and $\mu$ for $\Delta\phi= \pi$. The black line shows the outlines of $Q=-1$ and $Q=1$ regions. b) The gap $E_{gap}$ of a JJ with periodic boundaries as a function of $h_J$ and $\mu$ for $\Delta\phi= \pi$. c) The topological number $W$ as a function of $h_J$ and $\Delta\phi$ for $\mu=1t$. d) The gap $E_{gap}$ of a JJ with periodic boundaries as a function of $h_J$ and $\Delta\phi$ for $\mu=1t$. We used the parameters: $W_J = 5a_c$, $W_L = W_R = 20a_c$. e) and f) Modified JJ structure and $|\psi|^2$ of the lowest eigenvalue corresponding to MBS. The phases of four superconducting regions are set to: $\phi_1=\phi_2=\pi/2$, $\phi_3=\phi_4 = -\pi/2$. Dimensions used: $W_U = W_D = W_C = 50a_c$, $W_S = 600a_c$, $W_J = 5a_c$. Other parameters used: $h_J=0.4t$, $h_{SC} = 0$, $\mu = 2t$, $\Delta_0 = 0.3t$, $\Delta^\prime_0 = 0.06t$, $\theta$ = $\arctan(1/9) \approx \pi/9$.  }
    \label{fig:d+id}
\end{figure}

\textit{Planar JJ with $d+id^\prime$ pairing.} ~We consider $d+id^\prime$ pairing in Fig.~\ref{fig:d+id}. The phase diagram for $Q$ and $W$ as a function of $\mu$ and $h_J$ in Fig.~\ref{fig:d+id}a) looks somewhat similar to the case of $d$-wave pairing in Fig.~\ref{fig:MBS}d). The $id^\prime$ component of the pairing potential breaks the time-reversal symmetry and removes the gapless states in the bulk of 2DEG. At the same time, the topology of the bulk of the proximitized 2DEG corresponds to the even Chern number and results in the chiral edge modes~\cite{Sato2010} in the system in Fig.~\ref{fig:MBS}a). A finite size effect associated with the gapless excitations of the chiral edge modes of 2DEG results in gap closings and changes in $W$ shown in Fig.~\ref{fig:d+id}a), where a periodic boundary along the $y$ axis in Fig.~\ref{fig:MBS}a) has been used. In Fig.~\ref{fig:d+id}c), we show the phase diagram for $W$ as a function of $\phi$ and $h_J$, where one can identify a diamond-shaped region with $Q=-1$. Results in Figs.~\ref{fig:d+id}a) and c) suggest that MBS will coexist with the chiral edge modes in system in Fig.~\ref{fig:MBS}a). If not separated spatially, the chiral edge modes can hybridize with MBS. In Figs.~\ref{fig:d+id}b) and d), we study the gap inside of the planar JJ in Fig.~\ref{fig:MBS}a) with the periodic boundary along the $y$ axis where the outer edge modes have been removed. With the periodic condition applied, the only possible chiral edge modes are along the JJ which would result in gapless states. Since the JJ is taken to be narrow, it is possible for the chiral modes to have a hybridization. Figures~\ref{fig:d+id}b) and d) show that the planar JJ with the Zeeman term can gap out the edge modes running along the JJ for a large range of parameters.  Using Figs.~\ref{fig:d+id}b) and d), we suggest a setup in Fig.~\ref{fig:d+id}e) with an angled JJ that allows to separate the edge modes from MBS by using four superconducting regions. Using parameters $h_J=0.4t$, $h_{SC} = 0$, $\mu = 2t$, $\Delta_0 = 0.3t$, $\Delta^\prime_0 = 0.06t$, and phases of four superconducting regions $\phi_1=\phi_2=\pi/2$, $\phi_3=\phi_4 = -\pi/2$ for setup in Fig.~\ref{fig:d+id}e), we are able to realize robust MBS, as shown in Fig.~\ref{fig:d+id}f) by plotting the $|\psi|^2$ of the lowest eigenvalue of the system.

\textit{Conclusions.} ~We have studied the planar JJ comprising a 2DEG with strong spin-orbit coupling and $d$-wave superconductors. The proximity effect can induce the superconducting pairing potential in 2DEG with the same symmetry as the host superconductor. Here, we have considered different types of $d$-wave superconductors with high critical temperature and large intrinsic gap. Apart from superconductors with pure $d$-wave pairing, we have also considered a twisted bilayer $d+is$, and $d+id^\prime$ superconductors realizable in mechanically exfoliated copper oxide heterostructures~\cite{Can2021}. 
In the case of $d$-wave pairing, we have demonstrated that the planar JJ can lead to MBS for a wide range of parameters, in analogy to realizations based on $s$-wave superconductors; however, the presence of gapless excitations in the bulk of 2DEG may hinder the quantum coherence. Nevertheless, we expect interesting manifestations in the superconducting diode effect~\cite{Tanaka2022}. In the case of $d+is$ pairing, we have demonstrated realizations of robust MBS for a wide range of parameters. 
In the case of $d+id^\prime$ pairing there are no gapless states in the bulk of 2DEG; however, the even Chern number associated with the bulk leads to the appearance of gapless chiral edge modes, which can hybridize with MBS. To realize MBS with $d+id^\prime$ pairing, we have proposed a modified JJ in which the chiral edge modes are gapped and do not hybridize with MBS. It would be interesting to consider generalizations of our ideas to 2DEG with cubic Rashba interactions~\cite{Alidoust2021,Luethi2023}.

The cuprate-based superconductors have shown critical temperatures of up to 133 K~\cite{Keimer2015}. 
One direct advantage of using cuprate-based superconductors is that MBS can exist at higher temperatures compared to realizations based on pure $s$-wave superconductors. Another advantage stems from much larger intrinsic gap which should result in a larger proximity-induced topological gap and better protection against disorder and thermal excitations~\cite{Court_2008,Wang2013,Yu2019}. Our results should help in identifying different platforms for realizations of robust and easily tunable MBS with better protection against decoherence.

\textit{Acknowledgment.} ~This work was supported by the U.S. Department of Energy, Office of Science, Basic Energy Sciences, under
Award No. DE-SC0021019.

\bibliography{main}

\begin{thebibliography}{52}%
\makeatletter
\providecommand \@ifxundefined [1]{%
 \@ifx{#1\undefined}
}%
\providecommand \@ifnum [1]{%
 \ifnum #1\expandafter \@firstoftwo
 \else \expandafter \@secondoftwo
 \fi
}%
\providecommand \@ifx [1]{%
 \ifx #1\expandafter \@firstoftwo
 \else \expandafter \@secondoftwo
 \fi
}%
\providecommand \natexlab [1]{#1}%
\providecommand \enquote  [1]{``#1''}%
\providecommand \bibnamefont  [1]{#1}%
\providecommand \bibfnamefont [1]{#1}%
\providecommand \citenamefont [1]{#1}%
\providecommand \href@noop [0]{\@secondoftwo}%
\providecommand \href [0]{\begingroup \@sanitize@url \@href}%
\providecommand \@href[1]{\@@startlink{#1}\@@href}%
\providecommand \@@href[1]{\endgroup#1\@@endlink}%
\providecommand \@sanitize@url [0]{\catcode `\\12\catcode `\$12\catcode `\&12\catcode `\#12\catcode `\^12\catcode `\_12\catcode `\%12\relax}%
\providecommand \@@startlink[1]{}%
\providecommand \@@endlink[0]{}%
\providecommand \url  [0]{\begingroup\@sanitize@url \@url }%
\providecommand \@url [1]{\endgroup\@href {#1}{\urlprefix }}%
\providecommand \urlprefix  [0]{URL }%
\providecommand \Eprint [0]{\href }%
\providecommand \doibase [0]{https://doi.org/}%
\providecommand \selectlanguage [0]{\@gobble}%
\providecommand \bibinfo  [0]{\@secondoftwo}%
\providecommand \bibfield  [0]{\@secondoftwo}%
\providecommand \translation [1]{[#1]}%
\providecommand \BibitemOpen [0]{}%
\providecommand \bibitemStop [0]{}%
\providecommand \bibitemNoStop [0]{.\EOS\space}%
\providecommand \EOS [0]{\spacefactor3000\relax}%
\providecommand \BibitemShut  [1]{\csname bibitem#1\endcsname}%
\let\auto@bib@innerbib\@empty
\bibitem [{\citenamefont {Nayak}\ \emph {et~al.}(2008)\citenamefont {Nayak}, \citenamefont {Simon}, \citenamefont {Stern}, \citenamefont {Freedman},\ and\ \citenamefont {Das~Sarma}}]{Nayak2008}%
  \BibitemOpen
  \bibfield  {author} {\bibinfo {author} {\bibfnamefont {C.}~\bibnamefont {Nayak}}, \bibinfo {author} {\bibfnamefont {S.~H.}\ \bibnamefont {Simon}}, \bibinfo {author} {\bibfnamefont {A.}~\bibnamefont {Stern}}, \bibinfo {author} {\bibfnamefont {M.}~\bibnamefont {Freedman}},\ and\ \bibinfo {author} {\bibfnamefont {S.}~\bibnamefont {Das~Sarma}},\ }\bibfield  {title} {\bibinfo {title} {Non-abelian anyons and topological quantum computation},\ }\href {https://link.aps.org/doi/10.1103/RevModPhys.80.1083} {\bibfield  {journal} {\bibinfo  {journal} {Rev. Mod. Phys.}\ }\textbf {\bibinfo {volume} {80}},\ \bibinfo {pages} {1083} (\bibinfo {year} {2008})}\BibitemShut {NoStop}%
\bibitem [{\citenamefont {Kitaev}(2003)}]{Kitaev2003}%
  \BibitemOpen
  \bibfield  {author} {\bibinfo {author} {\bibfnamefont {A.}~\bibnamefont {Kitaev}},\ }\bibfield  {title} {\bibinfo {title} {Fault-tolerant quantum computation by anyons},\ }\href {https://doi.org/10.1016/s0003-4916(02)00018-0} {\bibfield  {journal} {\bibinfo  {journal} {Ann. Phys.}\ }\textbf {\bibinfo {volume} {303}},\ \bibinfo {pages} {2} (\bibinfo {year} {2003})}\BibitemShut {NoStop}%
\bibitem [{\citenamefont {Alicea}(2012)}]{Alicea2012}%
  \BibitemOpen
  \bibfield  {author} {\bibinfo {author} {\bibfnamefont {J.}~\bibnamefont {Alicea}},\ }\bibfield  {title} {\bibinfo {title} {{New directions in the pursuit of Majorana fermions in solid state systems}},\ }\href {https://doi.org/10.1088/0034-4885/75/7/076501} {\bibfield  {journal} {\bibinfo  {journal} {Rep. Prog. Phys.}\ }\textbf {\bibinfo {volume} {75}},\ \bibinfo {pages} {076501} (\bibinfo {year} {2012})}\BibitemShut {NoStop}%
\bibitem [{\citenamefont {Beenakker}(2013)}]{Beenakker2013}%
  \BibitemOpen
  \bibfield  {author} {\bibinfo {author} {\bibfnamefont {C.}~\bibnamefont {Beenakker}},\ }\bibfield  {title} {\bibinfo {title} {{Search for Majorana Fermions in Superconductors}},\ }\href {https://doi.org/10.1146/annurev-conmatphys-030212-184337} {\bibfield  {journal} {\bibinfo  {journal} {Annu. Rev. Condens. Matter Phys.}\ }\textbf {\bibinfo {volume} {4}},\ \bibinfo {pages} {113} (\bibinfo {year} {2013})}\BibitemShut {NoStop}%
\bibitem [{\citenamefont {Elliott}\ and\ \citenamefont {Franz}(2015)}]{Elliott2015}%
  \BibitemOpen
  \bibfield  {author} {\bibinfo {author} {\bibfnamefont {S.~R.}\ \bibnamefont {Elliott}}\ and\ \bibinfo {author} {\bibfnamefont {M.}~\bibnamefont {Franz}},\ }\bibfield  {title} {\bibinfo {title} {{Colloquium: Majorana fermions in nuclear, particle, and solid-state physics}},\ }\href {https://doi.org/10.1103/RevModPhys.87.137} {\bibfield  {journal} {\bibinfo  {journal} {Rev. Mod. Phys.}\ }\textbf {\bibinfo {volume} {87}},\ \bibinfo {pages} {137} (\bibinfo {year} {2015})}\BibitemShut {NoStop}%
\bibitem [{\citenamefont {Fu}\ and\ \citenamefont {Kane}(2007)}]{FuKane}%
  \BibitemOpen
  \bibfield  {author} {\bibinfo {author} {\bibfnamefont {L.}~\bibnamefont {Fu}}\ and\ \bibinfo {author} {\bibfnamefont {C.~L.}\ \bibnamefont {Kane}},\ }\bibfield  {title} {\bibinfo {title} {Topological insulators with inversion symmetry},\ }\href {https://doi.org/10.1103/PhysRevB.76.045302} {\bibfield  {journal} {\bibinfo  {journal} {Phys. Rev. B}\ }\textbf {\bibinfo {volume} {76}},\ \bibinfo {pages} {045302} (\bibinfo {year} {2007})}\BibitemShut {NoStop}%
\bibitem [{\citenamefont {Eschrig}(2019)}]{Eschrig2019}%
  \BibitemOpen
  \bibfield  {author} {\bibinfo {author} {\bibfnamefont {M.}~\bibnamefont {Eschrig}},\ }\bibinfo {title} {Phase-sensitive interface and proximity effects in superconducting spintronics},\ in\ \href {https://www.taylorfrancis.com/chapters/edit/10.1201/9780429423079-16/phase-sensitive-interface-proximity-effects-superconducting-spintronics-matthias-eschrig} {\emph {\bibinfo {booktitle} {Spintronics Handbook: Spin Transport and Magnetism}}},\ \bibinfo {editor} {edited by\ \bibinfo {editor} {\bibfnamefont {E.~Y.}\ \bibnamefont {Tsymbal}}\ and\ \bibinfo {editor} {\bibfnamefont {I.}~\bibnamefont {Zutic}}}\ (\bibinfo  {publisher} {Taylor \& Francis},\ \bibinfo {address} {New York},\ \bibinfo {year} {2019})\ \bibinfo {edition} {2nd}\ ed.\BibitemShut {Stop}%
\bibitem [{\citenamefont {Setiawan}\ \emph {et~al.}(2019{\natexlab{a}})\citenamefont {Setiawan}, \citenamefont {Wu},\ and\ \citenamefont {Levin}}]{proximitylevin}%
  \BibitemOpen
  \bibfield  {author} {\bibinfo {author} {\bibfnamefont {F.}~\bibnamefont {Setiawan}}, \bibinfo {author} {\bibfnamefont {C.-T.}\ \bibnamefont {Wu}},\ and\ \bibinfo {author} {\bibfnamefont {K.}~\bibnamefont {Levin}},\ }\bibfield  {title} {\bibinfo {title} {Full proximity treatment of topological superconductors in josephson-junction architectures},\ }\href {https://doi.org/10.1103/PhysRevB.99.174511} {\bibfield  {journal} {\bibinfo  {journal} {Phys. Rev. B}\ }\textbf {\bibinfo {volume} {99}},\ \bibinfo {pages} {174511} (\bibinfo {year} {2019}{\natexlab{a}})}\BibitemShut {NoStop}%
\bibitem [{\citenamefont {{\v{Z}}uti{\'{c}}}\ \emph {et~al.}(2019)\citenamefont {{\v{Z}}uti{\'{c}}}, \citenamefont {Matos-Abiague}, \citenamefont {Scharf}, \citenamefont {Dery},\ and\ \citenamefont {Belashchenko}}]{Zutic2019}%
  \BibitemOpen
  \bibfield  {author} {\bibinfo {author} {\bibfnamefont {I.}~\bibnamefont {{\v{Z}}uti{\'{c}}}}, \bibinfo {author} {\bibfnamefont {A.}~\bibnamefont {Matos-Abiague}}, \bibinfo {author} {\bibfnamefont {B.}~\bibnamefont {Scharf}}, \bibinfo {author} {\bibfnamefont {H.}~\bibnamefont {Dery}},\ and\ \bibinfo {author} {\bibfnamefont {K.}~\bibnamefont {Belashchenko}},\ }\bibfield  {title} {\bibinfo {title} {Proximitized materials},\ }\href {https://doi.org/10.1016/j.mattod.2018.05.003} {\bibfield  {journal} {\bibinfo  {journal} {Mater. Today}\ }\textbf {\bibinfo {volume} {22}},\ \bibinfo {pages} {85} (\bibinfo {year} {2019})}\BibitemShut {NoStop}%
\bibitem [{\citenamefont {G\"{u}ng\"{o}rd\"{u}}\ and\ \citenamefont {Kovalev}(2022)}]{Gngrd2022}%
  \BibitemOpen
  \bibfield  {author} {\bibinfo {author} {\bibfnamefont {U.}~\bibnamefont {G\"{u}ng\"{o}rd\"{u}}}\ and\ \bibinfo {author} {\bibfnamefont {A.~A.}\ \bibnamefont {Kovalev}},\ }\bibfield  {title} {\bibinfo {title} {Majorana bound states with chiral magnetic textures},\ }\href {https://doi.org/10.1063/5.0097008} {\bibfield  {journal} {\bibinfo  {journal} {J. Appl. Phys.}\ }\textbf {\bibinfo {volume} {132}},\ \bibinfo {pages} {041101} (\bibinfo {year} {2022})}\BibitemShut {NoStop}%
\bibitem [{\citenamefont {Takei}\ \emph {et~al.}(2013)\citenamefont {Takei}, \citenamefont {Fregoso}, \citenamefont {Hui}, \citenamefont {Lobos},\ and\ \citenamefont {Das~Sarma}}]{Takei2013}%
  \BibitemOpen
  \bibfield  {author} {\bibinfo {author} {\bibfnamefont {S.}~\bibnamefont {Takei}}, \bibinfo {author} {\bibfnamefont {B.~M.}\ \bibnamefont {Fregoso}}, \bibinfo {author} {\bibfnamefont {H.-Y.}\ \bibnamefont {Hui}}, \bibinfo {author} {\bibfnamefont {A.~M.}\ \bibnamefont {Lobos}},\ and\ \bibinfo {author} {\bibfnamefont {S.}~\bibnamefont {Das~Sarma}},\ }\bibfield  {title} {\bibinfo {title} {Soft superconducting gap in semiconductor majorana nanowires},\ }\href {https://doi.org/10.1103/PhysRevLett.110.186803} {\bibfield  {journal} {\bibinfo  {journal} {Phys. Rev. Lett.}\ }\textbf {\bibinfo {volume} {110}},\ \bibinfo {pages} {186803} (\bibinfo {year} {2013})}\BibitemShut {NoStop}%
\bibitem [{\citenamefont {Zhang}\ \emph {et~al.}(2019)\citenamefont {Zhang}, \citenamefont {Liu}, \citenamefont {Wimmer},\ and\ \citenamefont {Kouwenhoven}}]{Zhang2019}%
  \BibitemOpen
  \bibfield  {author} {\bibinfo {author} {\bibfnamefont {H.}~\bibnamefont {Zhang}}, \bibinfo {author} {\bibfnamefont {D.~E.}\ \bibnamefont {Liu}}, \bibinfo {author} {\bibfnamefont {M.}~\bibnamefont {Wimmer}},\ and\ \bibinfo {author} {\bibfnamefont {L.~P.}\ \bibnamefont {Kouwenhoven}},\ }\bibfield  {title} {\bibinfo {title} {{Next steps of quantum transport in Majorana nanowire devices}},\ }\href {https://doi.org/10.1038/s41467-019-13133-1} {\bibfield  {journal} {\bibinfo  {journal} {Nat. Commun.}\ }\textbf {\bibinfo {volume} {10}},\ \bibinfo {pages} {5128} (\bibinfo {year} {2019})}\BibitemShut {NoStop}%
\bibitem [{\citenamefont {Rainis}\ and\ \citenamefont {Loss}(2012)}]{Rainis2012}%
  \BibitemOpen
  \bibfield  {author} {\bibinfo {author} {\bibfnamefont {D.}~\bibnamefont {Rainis}}\ and\ \bibinfo {author} {\bibfnamefont {D.}~\bibnamefont {Loss}},\ }\bibfield  {title} {\bibinfo {title} {Majorana qubit decoherence by quasiparticle poisoning},\ }\href {https://doi.org/10.1103/PhysRevB.85.174533} {\bibfield  {journal} {\bibinfo  {journal} {Phys. Rev. B}\ }\textbf {\bibinfo {volume} {85}},\ \bibinfo {pages} {174533} (\bibinfo {year} {2012})}\BibitemShut {NoStop}%
\bibitem [{\citenamefont {Stanescu}\ \emph {et~al.}(2014)\citenamefont {Stanescu}, \citenamefont {Lutchyn},\ and\ \citenamefont {Das~Sarma}}]{Stanescu2014}%
  \BibitemOpen
  \bibfield  {author} {\bibinfo {author} {\bibfnamefont {T.~D.}\ \bibnamefont {Stanescu}}, \bibinfo {author} {\bibfnamefont {R.~M.}\ \bibnamefont {Lutchyn}},\ and\ \bibinfo {author} {\bibfnamefont {S.}~\bibnamefont {Das~Sarma}},\ }\bibfield  {title} {\bibinfo {title} {Soft superconducting gap in semiconductor-based majorana nanowires},\ }\href {https://doi.org/10.1103/PhysRevB.90.085302} {\bibfield  {journal} {\bibinfo  {journal} {Phys. Rev. B}\ }\textbf {\bibinfo {volume} {90}},\ \bibinfo {pages} {085302} (\bibinfo {year} {2014})}\BibitemShut {NoStop}%
\bibitem [{\citenamefont {Pikulin}\ \emph {et~al.}()\citenamefont {Pikulin}, \citenamefont {van Heck}, \citenamefont {Karzig}, \citenamefont {Martinez}, \citenamefont {Nijholt}, \citenamefont {Laeven}, \citenamefont {Winkler}, \citenamefont {Watson}, \citenamefont {Heedt}, \citenamefont {Temurhan}, \citenamefont {Svidenko}, \citenamefont {Lutchyn}, \citenamefont {Thomas}, \citenamefont {de~Lange}, \citenamefont {Casparis},\ and\ \citenamefont {Nayak}}]{pikulin_protocol_2021}%
  \BibitemOpen
  \bibfield  {author} {\bibinfo {author} {\bibfnamefont {D.~I.}\ \bibnamefont {Pikulin}}, \bibinfo {author} {\bibfnamefont {B.}~\bibnamefont {van Heck}}, \bibinfo {author} {\bibfnamefont {T.}~\bibnamefont {Karzig}}, \bibinfo {author} {\bibfnamefont {E.~A.}\ \bibnamefont {Martinez}}, \bibinfo {author} {\bibfnamefont {B.}~\bibnamefont {Nijholt}}, \bibinfo {author} {\bibfnamefont {T.}~\bibnamefont {Laeven}}, \bibinfo {author} {\bibfnamefont {G.~W.}\ \bibnamefont {Winkler}}, \bibinfo {author} {\bibfnamefont {J.~D.}\ \bibnamefont {Watson}}, \bibinfo {author} {\bibfnamefont {S.}~\bibnamefont {Heedt}}, \bibinfo {author} {\bibfnamefont {M.}~\bibnamefont {Temurhan}}, \bibinfo {author} {\bibfnamefont {V.}~\bibnamefont {Svidenko}}, \bibinfo {author} {\bibfnamefont {R.~M.}\ \bibnamefont {Lutchyn}}, \bibinfo {author} {\bibfnamefont {M.}~\bibnamefont {Thomas}}, \bibinfo {author} {\bibfnamefont {G.}~\bibnamefont {de~Lange}}, \bibinfo {author} {\bibfnamefont {L.}~\bibnamefont {Casparis}},\ and\ \bibinfo {author}
  {\bibfnamefont {C.}~\bibnamefont {Nayak}},\ }\href {https://doi.org/10.48550/ARXIV.2103.12217} {\bibinfo {title} {Protocol to identify a topological superconducting phase in a three-terminal device}},\ \bibinfo {note} {arXiv.2103.12217, (2021)}\BibitemShut {NoStop}%
\bibitem [{\citenamefont {Aghaee}\ \emph {et~al.}(2023)\citenamefont {Aghaee}, \citenamefont {Akkala}, \citenamefont {Alam}, \citenamefont {Ali}, \citenamefont {Alcaraz~Ramirez}, \citenamefont {Andrzejczuk}, \citenamefont {Antipov}, \citenamefont {Aseev}, \citenamefont {Astafev}, \citenamefont {Bauer}, \citenamefont {Becker}, \citenamefont {Boddapati}, \citenamefont {Boekhout}, \citenamefont {Bommer}, \citenamefont {Bosma}, \citenamefont {Bourdet}, \citenamefont {Boutin}, \citenamefont {Caroff}, \citenamefont {Casparis}, \citenamefont {Cassidy}, \citenamefont {Chatoor}, \citenamefont {Christensen}, \citenamefont {Clay}, \citenamefont {Cole}, \citenamefont {Corsetti}, \citenamefont {Cui}, \citenamefont {Dalampiras}, \citenamefont {Dokania}, \citenamefont {de~Lange}, \citenamefont {de~Moor}, \citenamefont {Estrada Salda\~na}, \citenamefont {Fallahi}, \citenamefont {Fathabad}, \citenamefont {Gamble}, \citenamefont {Gardner}, \citenamefont {Govender}, \citenamefont {Griggio}, \citenamefont {Grigoryan},
  \citenamefont {Gronin}, \citenamefont {Gukelberger}, \citenamefont {Hansen}, \citenamefont {Heedt}, \citenamefont {Herranz~Zamorano}, \citenamefont {Ho}, \citenamefont {Holgaard}, \citenamefont {Ingerslev}, \citenamefont {Johansson}, \citenamefont {Jones}, \citenamefont {Kallaher}, \citenamefont {Karimi}, \citenamefont {Karzig}, \citenamefont {King}, \citenamefont {Kloster}, \citenamefont {Knapp}, \citenamefont {Kocon}, \citenamefont {Koski}, \citenamefont {Kostamo}, \citenamefont {Krogstrup}, \citenamefont {Kumar}, \citenamefont {Laeven}, \citenamefont {Larsen}, \citenamefont {Li}, \citenamefont {Lindemann}, \citenamefont {Love}, \citenamefont {Lutchyn}, \citenamefont {Madsen}, \citenamefont {Manfra}, \citenamefont {Markussen}, \citenamefont {Martinez}, \citenamefont {McNeil}, \citenamefont {Memisevic}, \citenamefont {Morgan}, \citenamefont {Mullally}, \citenamefont {Nayak}, \citenamefont {Nielsen}, \citenamefont {Nielsen}, \citenamefont {Nijholt}, \citenamefont {Nurmohamed}, \citenamefont {O'Farrell},
  \citenamefont {Otani}, \citenamefont {Pauka}, \citenamefont {Petersson}, \citenamefont {Petit}, \citenamefont {Pikulin}, \citenamefont {Preiss}, \citenamefont {Quintero-Perez}, \citenamefont {Rajpalke}, \citenamefont {Rasmussen}, \citenamefont {Razmadze}, \citenamefont {Reentila}, \citenamefont {Reilly}, \citenamefont {Rouse}, \citenamefont {Sadovskyy}, \citenamefont {Sainiemi}, \citenamefont {Schreppler}, \citenamefont {Sidorkin}, \citenamefont {Singh}, \citenamefont {Singh}, \citenamefont {Sinha}, \citenamefont {Sohr}, \citenamefont {Stankevi\ifmmode~\check{c}\else \v{c}\fi{}}, \citenamefont {Stek}, \citenamefont {Suominen}, \citenamefont {Suter}, \citenamefont {Svidenko}, \citenamefont {Teicher}, \citenamefont {Temuerhan}, \citenamefont {Thiyagarajah}, \citenamefont {Tholapi}, \citenamefont {Thomas}, \citenamefont {Toomey}, \citenamefont {Upadhyay}, \citenamefont {Urban}, \citenamefont {Vaitiek\ifmmode~\dot{e}\else \.{e}\fi{}nas}, \citenamefont {Van~Hoogdalem}, \citenamefont {Van~Woerkom}, \citenamefont
  {Viazmitinov}, \citenamefont {Vogel}, \citenamefont {Waddy}, \citenamefont {Watson}, \citenamefont {Weston}, \citenamefont {Winkler}, \citenamefont {Yang}, \citenamefont {Yau}, \citenamefont {Yi}, \citenamefont {Yucelen}, \citenamefont {Webster}, \citenamefont {Zeisel},\ and\ \citenamefont {Zhao}}]{aghaee_inas-hybrid_2022}%
  \BibitemOpen
  \bibfield  {author} {\bibinfo {author} {\bibfnamefont {M.}~\bibnamefont {Aghaee}}, \bibinfo {author} {\bibfnamefont {A.}~\bibnamefont {Akkala}}, \bibinfo {author} {\bibfnamefont {Z.}~\bibnamefont {Alam}}, \bibinfo {author} {\bibfnamefont {R.}~\bibnamefont {Ali}}, \bibinfo {author} {\bibfnamefont {A.}~\bibnamefont {Alcaraz~Ramirez}}, \bibinfo {author} {\bibfnamefont {M.}~\bibnamefont {Andrzejczuk}}, \bibinfo {author} {\bibfnamefont {A.~E.}\ \bibnamefont {Antipov}}, \bibinfo {author} {\bibfnamefont {P.}~\bibnamefont {Aseev}}, \bibinfo {author} {\bibfnamefont {M.}~\bibnamefont {Astafev}}, \bibinfo {author} {\bibfnamefont {B.}~\bibnamefont {Bauer}}, \bibinfo {author} {\bibfnamefont {J.}~\bibnamefont {Becker}}, \bibinfo {author} {\bibfnamefont {S.}~\bibnamefont {Boddapati}}, \bibinfo {author} {\bibfnamefont {F.}~\bibnamefont {Boekhout}}, \bibinfo {author} {\bibfnamefont {J.}~\bibnamefont {Bommer}}, \bibinfo {author} {\bibfnamefont {T.}~\bibnamefont {Bosma}}, \bibinfo {author} {\bibfnamefont {L.}~\bibnamefont
  {Bourdet}}, \bibinfo {author} {\bibfnamefont {S.}~\bibnamefont {Boutin}}, \bibinfo {author} {\bibfnamefont {P.}~\bibnamefont {Caroff}}, \bibinfo {author} {\bibfnamefont {L.}~\bibnamefont {Casparis}}, \bibinfo {author} {\bibfnamefont {M.}~\bibnamefont {Cassidy}}, \bibinfo {author} {\bibfnamefont {S.}~\bibnamefont {Chatoor}}, \bibinfo {author} {\bibfnamefont {A.~W.}\ \bibnamefont {Christensen}}, \bibinfo {author} {\bibfnamefont {N.}~\bibnamefont {Clay}}, \bibinfo {author} {\bibfnamefont {W.~S.}\ \bibnamefont {Cole}}, \bibinfo {author} {\bibfnamefont {F.}~\bibnamefont {Corsetti}}, \bibinfo {author} {\bibfnamefont {A.}~\bibnamefont {Cui}}, \bibinfo {author} {\bibfnamefont {P.}~\bibnamefont {Dalampiras}}, \bibinfo {author} {\bibfnamefont {A.}~\bibnamefont {Dokania}}, \bibinfo {author} {\bibfnamefont {G.}~\bibnamefont {de~Lange}}, \bibinfo {author} {\bibfnamefont {M.}~\bibnamefont {de~Moor}}, \bibinfo {author} {\bibfnamefont {J.~C.}\ \bibnamefont {Estrada Salda\~na}}, \bibinfo {author} {\bibfnamefont
  {S.}~\bibnamefont {Fallahi}}, \bibinfo {author} {\bibfnamefont {Z.~H.}\ \bibnamefont {Fathabad}}, \bibinfo {author} {\bibfnamefont {J.}~\bibnamefont {Gamble}}, \bibinfo {author} {\bibfnamefont {G.}~\bibnamefont {Gardner}}, \bibinfo {author} {\bibfnamefont {D.}~\bibnamefont {Govender}}, \bibinfo {author} {\bibfnamefont {F.}~\bibnamefont {Griggio}}, \bibinfo {author} {\bibfnamefont {R.}~\bibnamefont {Grigoryan}}, \bibinfo {author} {\bibfnamefont {S.}~\bibnamefont {Gronin}}, \bibinfo {author} {\bibfnamefont {J.}~\bibnamefont {Gukelberger}}, \bibinfo {author} {\bibfnamefont {E.~B.}\ \bibnamefont {Hansen}}, \bibinfo {author} {\bibfnamefont {S.}~\bibnamefont {Heedt}}, \bibinfo {author} {\bibfnamefont {J.}~\bibnamefont {Herranz~Zamorano}}, \bibinfo {author} {\bibfnamefont {S.}~\bibnamefont {Ho}}, \bibinfo {author} {\bibfnamefont {U.~L.}\ \bibnamefont {Holgaard}}, \bibinfo {author} {\bibfnamefont {H.}~\bibnamefont {Ingerslev}}, \bibinfo {author} {\bibfnamefont {L.}~\bibnamefont {Johansson}}, \bibinfo {author}
  {\bibfnamefont {J.}~\bibnamefont {Jones}}, \bibinfo {author} {\bibfnamefont {R.}~\bibnamefont {Kallaher}}, \bibinfo {author} {\bibfnamefont {F.}~\bibnamefont {Karimi}}, \bibinfo {author} {\bibfnamefont {T.}~\bibnamefont {Karzig}}, \bibinfo {author} {\bibfnamefont {C.}~\bibnamefont {King}}, \bibinfo {author} {\bibfnamefont {M.~E.}\ \bibnamefont {Kloster}}, \bibinfo {author} {\bibfnamefont {C.}~\bibnamefont {Knapp}}, \bibinfo {author} {\bibfnamefont {D.}~\bibnamefont {Kocon}}, \bibinfo {author} {\bibfnamefont {J.}~\bibnamefont {Koski}}, \bibinfo {author} {\bibfnamefont {P.}~\bibnamefont {Kostamo}}, \bibinfo {author} {\bibfnamefont {P.}~\bibnamefont {Krogstrup}}, \bibinfo {author} {\bibfnamefont {M.}~\bibnamefont {Kumar}}, \bibinfo {author} {\bibfnamefont {T.}~\bibnamefont {Laeven}}, \bibinfo {author} {\bibfnamefont {T.}~\bibnamefont {Larsen}}, \bibinfo {author} {\bibfnamefont {K.}~\bibnamefont {Li}}, \bibinfo {author} {\bibfnamefont {T.}~\bibnamefont {Lindemann}}, \bibinfo {author} {\bibfnamefont
  {J.}~\bibnamefont {Love}}, \bibinfo {author} {\bibfnamefont {R.}~\bibnamefont {Lutchyn}}, \bibinfo {author} {\bibfnamefont {M.~H.}\ \bibnamefont {Madsen}}, \bibinfo {author} {\bibfnamefont {M.}~\bibnamefont {Manfra}}, \bibinfo {author} {\bibfnamefont {S.}~\bibnamefont {Markussen}}, \bibinfo {author} {\bibfnamefont {E.}~\bibnamefont {Martinez}}, \bibinfo {author} {\bibfnamefont {R.}~\bibnamefont {McNeil}}, \bibinfo {author} {\bibfnamefont {E.}~\bibnamefont {Memisevic}}, \bibinfo {author} {\bibfnamefont {T.}~\bibnamefont {Morgan}}, \bibinfo {author} {\bibfnamefont {A.}~\bibnamefont {Mullally}}, \bibinfo {author} {\bibfnamefont {C.}~\bibnamefont {Nayak}}, \bibinfo {author} {\bibfnamefont {J.}~\bibnamefont {Nielsen}}, \bibinfo {author} {\bibfnamefont {W.~H.~P.}\ \bibnamefont {Nielsen}}, \bibinfo {author} {\bibfnamefont {B.}~\bibnamefont {Nijholt}}, \bibinfo {author} {\bibfnamefont {A.}~\bibnamefont {Nurmohamed}}, \bibinfo {author} {\bibfnamefont {E.}~\bibnamefont {O'Farrell}}, \bibinfo {author} {\bibfnamefont
  {K.}~\bibnamefont {Otani}}, \bibinfo {author} {\bibfnamefont {S.}~\bibnamefont {Pauka}}, \bibinfo {author} {\bibfnamefont {K.}~\bibnamefont {Petersson}}, \bibinfo {author} {\bibfnamefont {L.}~\bibnamefont {Petit}}, \bibinfo {author} {\bibfnamefont {D.~I.}\ \bibnamefont {Pikulin}}, \bibinfo {author} {\bibfnamefont {F.}~\bibnamefont {Preiss}}, \bibinfo {author} {\bibfnamefont {M.}~\bibnamefont {Quintero-Perez}}, \bibinfo {author} {\bibfnamefont {M.}~\bibnamefont {Rajpalke}}, \bibinfo {author} {\bibfnamefont {K.}~\bibnamefont {Rasmussen}}, \bibinfo {author} {\bibfnamefont {D.}~\bibnamefont {Razmadze}}, \bibinfo {author} {\bibfnamefont {O.}~\bibnamefont {Reentila}}, \bibinfo {author} {\bibfnamefont {D.}~\bibnamefont {Reilly}}, \bibinfo {author} {\bibfnamefont {R.}~\bibnamefont {Rouse}}, \bibinfo {author} {\bibfnamefont {I.}~\bibnamefont {Sadovskyy}}, \bibinfo {author} {\bibfnamefont {L.}~\bibnamefont {Sainiemi}}, \bibinfo {author} {\bibfnamefont {S.}~\bibnamefont {Schreppler}}, \bibinfo {author} {\bibfnamefont
  {V.}~\bibnamefont {Sidorkin}}, \bibinfo {author} {\bibfnamefont {A.}~\bibnamefont {Singh}}, \bibinfo {author} {\bibfnamefont {S.}~\bibnamefont {Singh}}, \bibinfo {author} {\bibfnamefont {S.}~\bibnamefont {Sinha}}, \bibinfo {author} {\bibfnamefont {P.}~\bibnamefont {Sohr}}, \bibinfo {author} {\bibfnamefont {T.~c.~v.}\ \bibnamefont {Stankevi\ifmmode~\check{c}\else \v{c}\fi{}}}, \bibinfo {author} {\bibfnamefont {L.}~\bibnamefont {Stek}}, \bibinfo {author} {\bibfnamefont {H.}~\bibnamefont {Suominen}}, \bibinfo {author} {\bibfnamefont {J.}~\bibnamefont {Suter}}, \bibinfo {author} {\bibfnamefont {V.}~\bibnamefont {Svidenko}}, \bibinfo {author} {\bibfnamefont {S.}~\bibnamefont {Teicher}}, \bibinfo {author} {\bibfnamefont {M.}~\bibnamefont {Temuerhan}}, \bibinfo {author} {\bibfnamefont {N.}~\bibnamefont {Thiyagarajah}}, \bibinfo {author} {\bibfnamefont {R.}~\bibnamefont {Tholapi}}, \bibinfo {author} {\bibfnamefont {M.}~\bibnamefont {Thomas}}, \bibinfo {author} {\bibfnamefont {E.}~\bibnamefont {Toomey}}, \bibinfo
  {author} {\bibfnamefont {S.}~\bibnamefont {Upadhyay}}, \bibinfo {author} {\bibfnamefont {I.}~\bibnamefont {Urban}}, \bibinfo {author} {\bibfnamefont {S.}~\bibnamefont {Vaitiek\ifmmode~\dot{e}\else \.{e}\fi{}nas}}, \bibinfo {author} {\bibfnamefont {K.}~\bibnamefont {Van~Hoogdalem}}, \bibinfo {author} {\bibfnamefont {D.}~\bibnamefont {Van~Woerkom}}, \bibinfo {author} {\bibfnamefont {D.~V.}\ \bibnamefont {Viazmitinov}}, \bibinfo {author} {\bibfnamefont {D.}~\bibnamefont {Vogel}}, \bibinfo {author} {\bibfnamefont {S.}~\bibnamefont {Waddy}}, \bibinfo {author} {\bibfnamefont {J.}~\bibnamefont {Watson}}, \bibinfo {author} {\bibfnamefont {J.}~\bibnamefont {Weston}}, \bibinfo {author} {\bibfnamefont {G.~W.}\ \bibnamefont {Winkler}}, \bibinfo {author} {\bibfnamefont {C.~K.}\ \bibnamefont {Yang}}, \bibinfo {author} {\bibfnamefont {S.}~\bibnamefont {Yau}}, \bibinfo {author} {\bibfnamefont {D.}~\bibnamefont {Yi}}, \bibinfo {author} {\bibfnamefont {E.}~\bibnamefont {Yucelen}}, \bibinfo {author} {\bibfnamefont
  {A.}~\bibnamefont {Webster}}, \bibinfo {author} {\bibfnamefont {R.}~\bibnamefont {Zeisel}},\ and\ \bibinfo {author} {\bibfnamefont {R.}~\bibnamefont {Zhao}} (\bibinfo {collaboration} {Microsoft Quantum}),\ }\bibfield  {title} {\bibinfo {title} {Inas-al hybrid devices passing the topological gap protocol},\ }\href {https://doi.org/10.1103/PhysRevB.107.245423} {\bibfield  {journal} {\bibinfo  {journal} {Phys. Rev. B}\ }\textbf {\bibinfo {volume} {107}},\ \bibinfo {pages} {245423} (\bibinfo {year} {2023})}\BibitemShut {NoStop}%
\bibitem [{\citenamefont {Huo}\ \emph {et~al.}(2023)\citenamefont {Huo}, \citenamefont {Xia}, \citenamefont {Li}, \citenamefont {Zhang}, \citenamefont {Wang}, \citenamefont {Pan}, \citenamefont {Liu}, \citenamefont {Liu}, \citenamefont {Wang}, \citenamefont {Gao}, \citenamefont {Zhao}, \citenamefont {Li}, \citenamefont {Ying}, \citenamefont {Shang},\ and\ \citenamefont {Zhang}}]{Huo_2023}%
  \BibitemOpen
  \bibfield  {author} {\bibinfo {author} {\bibfnamefont {J.}~\bibnamefont {Huo}}, \bibinfo {author} {\bibfnamefont {Z.}~\bibnamefont {Xia}}, \bibinfo {author} {\bibfnamefont {Z.}~\bibnamefont {Li}}, \bibinfo {author} {\bibfnamefont {S.}~\bibnamefont {Zhang}}, \bibinfo {author} {\bibfnamefont {Y.}~\bibnamefont {Wang}}, \bibinfo {author} {\bibfnamefont {D.}~\bibnamefont {Pan}}, \bibinfo {author} {\bibfnamefont {Q.}~\bibnamefont {Liu}}, \bibinfo {author} {\bibfnamefont {Y.}~\bibnamefont {Liu}}, \bibinfo {author} {\bibfnamefont {Z.}~\bibnamefont {Wang}}, \bibinfo {author} {\bibfnamefont {Y.}~\bibnamefont {Gao}}, \bibinfo {author} {\bibfnamefont {J.}~\bibnamefont {Zhao}}, \bibinfo {author} {\bibfnamefont {T.}~\bibnamefont {Li}}, \bibinfo {author} {\bibfnamefont {J.}~\bibnamefont {Ying}}, \bibinfo {author} {\bibfnamefont {R.}~\bibnamefont {Shang}},\ and\ \bibinfo {author} {\bibfnamefont {H.}~\bibnamefont {Zhang}},\ }\bibfield  {title} {\bibinfo {title} {Gatemon qubit based on a thin inas-al hybrid nanowire},\
  }\href {https://doi.org/10.1088/0256-307X/40/4/047302} {\bibfield  {journal} {\bibinfo  {journal} {Chinese Phys. Lett.}\ }\textbf {\bibinfo {volume} {40}},\ \bibinfo {pages} {047302} (\bibinfo {year} {2023})}\BibitemShut {NoStop}%
\bibitem [{\citenamefont {Yanase}\ \emph {et~al.}(2003)\citenamefont {Yanase}, \citenamefont {Jujo}, \citenamefont {Nomura}, \citenamefont {Ikeda}, \citenamefont {Hotta},\ and\ \citenamefont {Yamada}}]{Yanase2003}%
  \BibitemOpen
  \bibfield  {author} {\bibinfo {author} {\bibfnamefont {Y.}~\bibnamefont {Yanase}}, \bibinfo {author} {\bibfnamefont {T.}~\bibnamefont {Jujo}}, \bibinfo {author} {\bibfnamefont {T.}~\bibnamefont {Nomura}}, \bibinfo {author} {\bibfnamefont {H.}~\bibnamefont {Ikeda}}, \bibinfo {author} {\bibfnamefont {T.}~\bibnamefont {Hotta}},\ and\ \bibinfo {author} {\bibfnamefont {K.}~\bibnamefont {Yamada}},\ }\bibfield  {title} {\bibinfo {title} {Theory of superconductivity in strongly correlated electron systems},\ }\href {https://doi.org/10.1016/j.physrep.2003.07.002} {\bibfield  {journal} {\bibinfo  {journal} {Phys. Rep.}\ }\textbf {\bibinfo {volume} {387}},\ \bibinfo {pages} {1} (\bibinfo {year} {2003})}\BibitemShut {NoStop}%
\bibitem [{\citenamefont {Sato}\ and\ \citenamefont {Ando}(2017)}]{Sato2017}%
  \BibitemOpen
  \bibfield  {author} {\bibinfo {author} {\bibfnamefont {M.}~\bibnamefont {Sato}}\ and\ \bibinfo {author} {\bibfnamefont {Y.}~\bibnamefont {Ando}},\ }\bibfield  {title} {\bibinfo {title} {Topological superconductors: a review},\ }\href {https://doi.org/10.1088/1361-6633/aa6ac7} {\bibfield  {journal} {\bibinfo  {journal} {Rep. Prog. Phys.}\ }\textbf {\bibinfo {volume} {80}},\ \bibinfo {pages} {076501} (\bibinfo {year} {2017})}\BibitemShut {NoStop}%
\bibitem [{\citenamefont {Adagideli}\ \emph {et~al.}(1999)\citenamefont {Adagideli}, \citenamefont {Goldbart}, \citenamefont {Shnirman},\ and\ \citenamefont {Yazdani}}]{Adagideli1999}%
  \BibitemOpen
  \bibfield  {author} {\bibinfo {author} {\bibfnamefont {I.}~\bibnamefont {Adagideli}}, \bibinfo {author} {\bibfnamefont {P.~M.}\ \bibnamefont {Goldbart}}, \bibinfo {author} {\bibfnamefont {A.}~\bibnamefont {Shnirman}},\ and\ \bibinfo {author} {\bibfnamefont {A.}~\bibnamefont {Yazdani}},\ }\bibfield  {title} {\bibinfo {title} {Low-energy quasiparticle states near extended scatterers in $\mathit{d}$-wave superconductors and their connection with susy quantum mechanics},\ }\href {https://doi.org/10.1103/physrevlett.83.5571} {\bibfield  {journal} {\bibinfo  {journal} {Phys. Rev. Lett.}\ }\textbf {\bibinfo {volume} {83}},\ \bibinfo {pages} {5571} (\bibinfo {year} {1999})}\BibitemShut {NoStop}%
\bibitem [{\citenamefont {Chen}\ \emph {et~al.}(2001)\citenamefont {Chen}, \citenamefont {Biswas}, \citenamefont {\ifmmode \check{Z}\else \v{Z}\fi{}uti\ifmmode~\acute{c}\else \'{c}\fi{}}, \citenamefont {Wu}, \citenamefont {Ogale}, \citenamefont {Greene},\ and\ \citenamefont {Venkatesan}}]{Chen2001}%
  \BibitemOpen
  \bibfield  {author} {\bibinfo {author} {\bibfnamefont {Z.~Y.}\ \bibnamefont {Chen}}, \bibinfo {author} {\bibfnamefont {A.}~\bibnamefont {Biswas}}, \bibinfo {author} {\bibfnamefont {I.}~\bibnamefont {\ifmmode \check{Z}\else \v{Z}\fi{}uti\ifmmode~\acute{c}\else \'{c}\fi{}}}, \bibinfo {author} {\bibfnamefont {T.}~\bibnamefont {Wu}}, \bibinfo {author} {\bibfnamefont {S.~B.}\ \bibnamefont {Ogale}}, \bibinfo {author} {\bibfnamefont {R.~L.}\ \bibnamefont {Greene}},\ and\ \bibinfo {author} {\bibfnamefont {T.}~\bibnamefont {Venkatesan}},\ }\bibfield  {title} {\bibinfo {title} {Spin-polarized transport across a ${\mathrm{la}}_{0.7}{\mathrm{sr}}_{0.3}{\mathrm{mno}}_{3}/{\mathrm{yba}}_{2}{\mathrm{cu}}_{3}{\mathrm{o}}_{7\ensuremath{-}x}$ interface: Role of andreev bound states},\ }\href {https://doi.org/10.1103/PhysRevB.63.212508} {\bibfield  {journal} {\bibinfo  {journal} {Phys. Rev. B}\ }\textbf {\bibinfo {volume} {63}},\ \bibinfo {pages} {212508} (\bibinfo {year} {2001})}\BibitemShut {NoStop}%
\bibitem [{\citenamefont {Sengupta}\ \emph {et~al.}(2001)\citenamefont {Sengupta}, \citenamefont {\ifmmode \check{Z}\else \v{Z}\fi{}uti\ifmmode~\acute{c}\else \'{c}\fi{}}, \citenamefont {Kwon}, \citenamefont {Yakovenko},\ and\ \citenamefont {Das~Sarma}}]{Sengupta2001}%
  \BibitemOpen
  \bibfield  {author} {\bibinfo {author} {\bibfnamefont {K.}~\bibnamefont {Sengupta}}, \bibinfo {author} {\bibfnamefont {I.}~\bibnamefont {\ifmmode \check{Z}\else \v{Z}\fi{}uti\ifmmode~\acute{c}\else \'{c}\fi{}}}, \bibinfo {author} {\bibfnamefont {H.-J.}\ \bibnamefont {Kwon}}, \bibinfo {author} {\bibfnamefont {V.~M.}\ \bibnamefont {Yakovenko}},\ and\ \bibinfo {author} {\bibfnamefont {S.}~\bibnamefont {Das~Sarma}},\ }\bibfield  {title} {\bibinfo {title} {Midgap edge states and pairing symmetry of quasi-one-dimensional organic superconductors},\ }\href {https://doi.org/10.1103/PhysRevB.63.144531} {\bibfield  {journal} {\bibinfo  {journal} {Phys. Rev. B}\ }\textbf {\bibinfo {volume} {63}},\ \bibinfo {pages} {144531} (\bibinfo {year} {2001})}\BibitemShut {NoStop}%
\bibitem [{\citenamefont {Tanaka}\ \emph {et~al.}(2012)\citenamefont {Tanaka}, \citenamefont {Sato},\ and\ \citenamefont {Nagaosa}}]{Tanaka2012}%
  \BibitemOpen
  \bibfield  {author} {\bibinfo {author} {\bibfnamefont {Y.}~\bibnamefont {Tanaka}}, \bibinfo {author} {\bibfnamefont {M.}~\bibnamefont {Sato}},\ and\ \bibinfo {author} {\bibfnamefont {N.}~\bibnamefont {Nagaosa}},\ }\bibfield  {title} {\bibinfo {title} {Symmetry and topology in superconductors {\textendash}odd-frequency pairing and edge states{\textendash}},\ }\href {https://doi.org/10.1143/jpsj.81.011013} {\bibfield  {journal} {\bibinfo  {journal} {J. Phys. Soc. Jpn}\ }\textbf {\bibinfo {volume} {81}},\ \bibinfo {pages} {011013} (\bibinfo {year} {2012})}\BibitemShut {NoStop}%
\bibitem [{\citenamefont {Li}\ \emph {et~al.}(2015)\citenamefont {Li}, \citenamefont {Chan},\ and\ \citenamefont {Yao}}]{MZMTI}%
  \BibitemOpen
  \bibfield  {author} {\bibinfo {author} {\bibfnamefont {Z.-X.}\ \bibnamefont {Li}}, \bibinfo {author} {\bibfnamefont {C.}~\bibnamefont {Chan}},\ and\ \bibinfo {author} {\bibfnamefont {H.}~\bibnamefont {Yao}},\ }\bibfield  {title} {\bibinfo {title} {Realizing majorana zero modes by proximity effect between topological insulators and $d$-wave high-temperature superconductors},\ }\href {https://doi.org/10.1103/PhysRevB.91.235143} {\bibfield  {journal} {\bibinfo  {journal} {Phys. Rev. B}\ }\textbf {\bibinfo {volume} {91}},\ \bibinfo {pages} {235143} (\bibinfo {year} {2015})}\BibitemShut {NoStop}%
\bibitem [{\citenamefont {Ortiz}\ \emph {et~al.}(2018)\citenamefont {Ortiz}, \citenamefont {Varona}, \citenamefont {Viyuela},\ and\ \citenamefont {Martin-Delgado}}]{dwave2deg}%
  \BibitemOpen
  \bibfield  {author} {\bibinfo {author} {\bibfnamefont {L.}~\bibnamefont {Ortiz}}, \bibinfo {author} {\bibfnamefont {S.}~\bibnamefont {Varona}}, \bibinfo {author} {\bibfnamefont {O.}~\bibnamefont {Viyuela}},\ and\ \bibinfo {author} {\bibfnamefont {M.~A.}\ \bibnamefont {Martin-Delgado}},\ }\bibfield  {title} {\bibinfo {title} {Localization and oscillations of majorana fermions in a two-dimensional electron gas coupled with $d$-wave superconductors},\ }\href {https://doi.org/10.1103/PhysRevB.97.064501} {\bibfield  {journal} {\bibinfo  {journal} {Phys. Rev. B}\ }\textbf {\bibinfo {volume} {97}},\ \bibinfo {pages} {064501} (\bibinfo {year} {2018})}\BibitemShut {NoStop}%
\bibitem [{\citenamefont {Liu}\ \emph {et~al.}(2018)\citenamefont {Liu}, \citenamefont {He},\ and\ \citenamefont {Nori}}]{Liu2018}%
  \BibitemOpen
  \bibfield  {author} {\bibinfo {author} {\bibfnamefont {T.}~\bibnamefont {Liu}}, \bibinfo {author} {\bibfnamefont {J.~J.}\ \bibnamefont {He}},\ and\ \bibinfo {author} {\bibfnamefont {F.}~\bibnamefont {Nori}},\ }\bibfield  {title} {\bibinfo {title} {Majorana corner states in a two-dimensional magnetic topological insulator on a high-temperature superconductor},\ }\href {https://doi.org/10.1103/PhysRevB.98.245413} {\bibfield  {journal} {\bibinfo  {journal} {Phys. Rev. B}\ }\textbf {\bibinfo {volume} {98}},\ \bibinfo {pages} {245413} (\bibinfo {year} {2018})}\BibitemShut {NoStop}%
\bibitem [{\citenamefont {Mashkoori}\ and\ \citenamefont {Black-Schaffer}(2019)}]{magnetchain}%
  \BibitemOpen
  \bibfield  {author} {\bibinfo {author} {\bibfnamefont {M.}~\bibnamefont {Mashkoori}}\ and\ \bibinfo {author} {\bibfnamefont {A.}~\bibnamefont {Black-Schaffer}},\ }\bibfield  {title} {\bibinfo {title} {Majorana bound states in magnetic impurity chains: Effects of $d$-wave pairing},\ }\href {https://doi.org/10.1103/PhysRevB.99.024505} {\bibfield  {journal} {\bibinfo  {journal} {Phys. Rev. B}\ }\textbf {\bibinfo {volume} {99}},\ \bibinfo {pages} {024505} (\bibinfo {year} {2019})}\BibitemShut {NoStop}%
\bibitem [{\citenamefont {Yanase}\ \emph {et~al.}(2022)\citenamefont {Yanase}, \citenamefont {Daido}, \citenamefont {Takasan},\ and\ \citenamefont {Yoshida}}]{Yanase2022}%
  \BibitemOpen
  \bibfield  {author} {\bibinfo {author} {\bibfnamefont {Y.}~\bibnamefont {Yanase}}, \bibinfo {author} {\bibfnamefont {A.}~\bibnamefont {Daido}}, \bibinfo {author} {\bibfnamefont {K.}~\bibnamefont {Takasan}},\ and\ \bibinfo {author} {\bibfnamefont {T.}~\bibnamefont {Yoshida}},\ }\bibfield  {title} {\bibinfo {title} {Topological d-wave superconductivity in two dimensions},\ }\href {https://doi.org/10.1016/j.physe.2022.115143} {\bibfield  {journal} {\bibinfo  {journal} {J. Phys. E.}\ }\textbf {\bibinfo {volume} {140}},\ \bibinfo {pages} {115143} (\bibinfo {year} {2022})}\BibitemShut {NoStop}%
\bibitem [{\citenamefont {Crawford}\ \emph {et~al.}(2020)\citenamefont {Crawford}, \citenamefont {Mascot}, \citenamefont {Morr},\ and\ \citenamefont {Rachel}}]{Crawford2020}%
  \BibitemOpen
  \bibfield  {author} {\bibinfo {author} {\bibfnamefont {D.}~\bibnamefont {Crawford}}, \bibinfo {author} {\bibfnamefont {E.}~\bibnamefont {Mascot}}, \bibinfo {author} {\bibfnamefont {D.~K.}\ \bibnamefont {Morr}},\ and\ \bibinfo {author} {\bibfnamefont {S.}~\bibnamefont {Rachel}},\ }\bibfield  {title} {\bibinfo {title} {High-temperature majorana fermions in magnet-superconductor hybrid systems},\ }\href {https://doi.org/10.1103/PhysRevB.101.174510} {\bibfield  {journal} {\bibinfo  {journal} {Phys. Rev. B}\ }\textbf {\bibinfo {volume} {101}},\ \bibinfo {pages} {174510} (\bibinfo {year} {2020})}\BibitemShut {NoStop}%
\bibitem [{\citenamefont {Mercado}\ \emph {et~al.}(2022)\citenamefont {Mercado}, \citenamefont {Sahoo},\ and\ \citenamefont {Franz}}]{Mercado2022}%
  \BibitemOpen
  \bibfield  {author} {\bibinfo {author} {\bibfnamefont {A.}~\bibnamefont {Mercado}}, \bibinfo {author} {\bibfnamefont {S.}~\bibnamefont {Sahoo}},\ and\ \bibinfo {author} {\bibfnamefont {M.}~\bibnamefont {Franz}},\ }\bibfield  {title} {\bibinfo {title} {High-temperature majorana zero modes},\ }\href {https://doi.org/10.1103/PhysRevLett.128.137002} {\bibfield  {journal} {\bibinfo  {journal} {Phys. Rev. Lett.}\ }\textbf {\bibinfo {volume} {128}},\ \bibinfo {pages} {137002} (\bibinfo {year} {2022})}\BibitemShut {NoStop}%
\bibitem [{\citenamefont {Margalit}\ \emph {et~al.}(2022)\citenamefont {Margalit}, \citenamefont {Yan}, \citenamefont {Franz},\ and\ \citenamefont {Oreg}}]{Margalit2022}%
  \BibitemOpen
  \bibfield  {author} {\bibinfo {author} {\bibfnamefont {G.}~\bibnamefont {Margalit}}, \bibinfo {author} {\bibfnamefont {B.}~\bibnamefont {Yan}}, \bibinfo {author} {\bibfnamefont {M.}~\bibnamefont {Franz}},\ and\ \bibinfo {author} {\bibfnamefont {Y.}~\bibnamefont {Oreg}},\ }\bibfield  {title} {\bibinfo {title} {Chiral majorana modes via proximity to a twisted cuprate bilayer},\ }\href {https://doi.org/10.1103/PhysRevB.106.205424} {\bibfield  {journal} {\bibinfo  {journal} {Phys. Rev. B}\ }\textbf {\bibinfo {volume} {106}},\ \bibinfo {pages} {205424} (\bibinfo {year} {2022})}\BibitemShut {NoStop}%
\bibitem [{\citenamefont {Can}\ \emph {et~al.}(2021)\citenamefont {Can}, \citenamefont {Tummuru}, \citenamefont {Day}, \citenamefont {Elfimov}, \citenamefont {Damascelli},\ and\ \citenamefont {Franz}}]{Can2021}%
  \BibitemOpen
  \bibfield  {author} {\bibinfo {author} {\bibfnamefont {O.}~\bibnamefont {Can}}, \bibinfo {author} {\bibfnamefont {T.}~\bibnamefont {Tummuru}}, \bibinfo {author} {\bibfnamefont {R.~P.}\ \bibnamefont {Day}}, \bibinfo {author} {\bibfnamefont {I.}~\bibnamefont {Elfimov}}, \bibinfo {author} {\bibfnamefont {A.}~\bibnamefont {Damascelli}},\ and\ \bibinfo {author} {\bibfnamefont {M.}~\bibnamefont {Franz}},\ }\bibfield  {title} {\bibinfo {title} {High-temperature topological superconductivity in twisted double-layer copper oxides},\ }\href {https://doi.org/10.1038/s41567-020-01142-7} {\bibfield  {journal} {\bibinfo  {journal} {Nat. Phys.}\ }\textbf {\bibinfo {volume} {17}},\ \bibinfo {pages} {519} (\bibinfo {year} {2021})}\BibitemShut {NoStop}%
\bibitem [{\citenamefont {Hell}\ \emph {et~al.}(2017)\citenamefont {Hell}, \citenamefont {Leijnse},\ and\ \citenamefont {Flensberg}}]{JJ2017}%
  \BibitemOpen
  \bibfield  {author} {\bibinfo {author} {\bibfnamefont {M.}~\bibnamefont {Hell}}, \bibinfo {author} {\bibfnamefont {M.}~\bibnamefont {Leijnse}},\ and\ \bibinfo {author} {\bibfnamefont {K.}~\bibnamefont {Flensberg}},\ }\bibfield  {title} {\bibinfo {title} {Two-dimensional platform for networks of majorana bound states},\ }\href {https://doi.org/10.1103/PhysRevLett.118.107701} {\bibfield  {journal} {\bibinfo  {journal} {Phys. Rev. Lett.}\ }\textbf {\bibinfo {volume} {118}},\ \bibinfo {pages} {107701} (\bibinfo {year} {2017})}\BibitemShut {NoStop}%
\bibitem [{\citenamefont {Pientka}\ \emph {et~al.}(2017)\citenamefont {Pientka}, \citenamefont {Keselman}, \citenamefont {Berg}, \citenamefont {Yacoby}, \citenamefont {Stern},\ and\ \citenamefont {Halperin}}]{Halperin2017}%
  \BibitemOpen
  \bibfield  {author} {\bibinfo {author} {\bibfnamefont {F.}~\bibnamefont {Pientka}}, \bibinfo {author} {\bibfnamefont {A.}~\bibnamefont {Keselman}}, \bibinfo {author} {\bibfnamefont {E.}~\bibnamefont {Berg}}, \bibinfo {author} {\bibfnamefont {A.}~\bibnamefont {Yacoby}}, \bibinfo {author} {\bibfnamefont {A.}~\bibnamefont {Stern}},\ and\ \bibinfo {author} {\bibfnamefont {B.~I.}\ \bibnamefont {Halperin}},\ }\bibfield  {title} {\bibinfo {title} {Topological superconductivity in a planar josephson junction},\ }\href {https://doi.org/10.1103/PhysRevX.7.021032} {\bibfield  {journal} {\bibinfo  {journal} {Phys. Rev. X}\ }\textbf {\bibinfo {volume} {7}},\ \bibinfo {pages} {021032} (\bibinfo {year} {2017})}\BibitemShut {NoStop}%
\bibitem [{\citenamefont {Fornieri}\ \emph {et~al.}(2019)\citenamefont {Fornieri}, \citenamefont {Whiticar}, \citenamefont {Setiawan}, \citenamefont {Portol{\'{e}}s}, \citenamefont {Drachmann}, \citenamefont {Keselman}, \citenamefont {Gronin}, \citenamefont {Thomas}, \citenamefont {Wang}, \citenamefont {Kallaher}, \citenamefont {Gardner}, \citenamefont {Berg}, \citenamefont {Manfra}, \citenamefont {Stern}, \citenamefont {Marcus},\ and\ \citenamefont {Nichele}}]{Fornieri2019}%
  \BibitemOpen
  \bibfield  {author} {\bibinfo {author} {\bibfnamefont {A.}~\bibnamefont {Fornieri}}, \bibinfo {author} {\bibfnamefont {A.~M.}\ \bibnamefont {Whiticar}}, \bibinfo {author} {\bibfnamefont {F.}~\bibnamefont {Setiawan}}, \bibinfo {author} {\bibfnamefont {E.}~\bibnamefont {Portol{\'{e}}s}}, \bibinfo {author} {\bibfnamefont {A.~C.~C.}\ \bibnamefont {Drachmann}}, \bibinfo {author} {\bibfnamefont {A.}~\bibnamefont {Keselman}}, \bibinfo {author} {\bibfnamefont {S.}~\bibnamefont {Gronin}}, \bibinfo {author} {\bibfnamefont {C.}~\bibnamefont {Thomas}}, \bibinfo {author} {\bibfnamefont {T.}~\bibnamefont {Wang}}, \bibinfo {author} {\bibfnamefont {R.}~\bibnamefont {Kallaher}}, \bibinfo {author} {\bibfnamefont {G.~C.}\ \bibnamefont {Gardner}}, \bibinfo {author} {\bibfnamefont {E.}~\bibnamefont {Berg}}, \bibinfo {author} {\bibfnamefont {M.~J.}\ \bibnamefont {Manfra}}, \bibinfo {author} {\bibfnamefont {A.}~\bibnamefont {Stern}}, \bibinfo {author} {\bibfnamefont {C.~M.}\ \bibnamefont {Marcus}},\ and\ \bibinfo {author}
  {\bibfnamefont {F.}~\bibnamefont {Nichele}},\ }\bibfield  {title} {\bibinfo {title} {Evidence of topological superconductivity in planar josephson junctions},\ }\href {https://doi.org/10.1038/s41586-019-1068-8} {\bibfield  {journal} {\bibinfo  {journal} {Nature}\ }\textbf {\bibinfo {volume} {569}},\ \bibinfo {pages} {89} (\bibinfo {year} {2019})}\BibitemShut {NoStop}%
\bibitem [{\citenamefont {Setiawan}\ \emph {et~al.}(2019{\natexlab{b}})\citenamefont {Setiawan}, \citenamefont {Stern},\ and\ \citenamefont {Berg}}]{narrowJJ}%
  \BibitemOpen
  \bibfield  {author} {\bibinfo {author} {\bibfnamefont {F.}~\bibnamefont {Setiawan}}, \bibinfo {author} {\bibfnamefont {A.}~\bibnamefont {Stern}},\ and\ \bibinfo {author} {\bibfnamefont {E.}~\bibnamefont {Berg}},\ }\bibfield  {title} {\bibinfo {title} {Topological superconductivity in planar josephson junctions: Narrowing down to the nanowire limit},\ }\href {https://doi.org/10.1103/PhysRevB.99.220506} {\bibfield  {journal} {\bibinfo  {journal} {Phys. Rev. B}\ }\textbf {\bibinfo {volume} {99}},\ \bibinfo {pages} {220506} (\bibinfo {year} {2019}{\natexlab{b}})}\BibitemShut {NoStop}%
\bibitem [{\citenamefont {Zhou}\ \emph {et~al.}(2022)\citenamefont {Zhou}, \citenamefont {Dartiailh}, \citenamefont {Sardashti}, \citenamefont {Han}, \citenamefont {Matos-Abiague}, \citenamefont {Shabani},\ and\ \citenamefont {{\v{Z}}uti{\'{c}}}}]{Zhou2022}%
  \BibitemOpen
  \bibfield  {author} {\bibinfo {author} {\bibfnamefont {T.}~\bibnamefont {Zhou}}, \bibinfo {author} {\bibfnamefont {M.~C.}\ \bibnamefont {Dartiailh}}, \bibinfo {author} {\bibfnamefont {K.}~\bibnamefont {Sardashti}}, \bibinfo {author} {\bibfnamefont {J.~E.}\ \bibnamefont {Han}}, \bibinfo {author} {\bibfnamefont {A.}~\bibnamefont {Matos-Abiague}}, \bibinfo {author} {\bibfnamefont {J.}~\bibnamefont {Shabani}},\ and\ \bibinfo {author} {\bibfnamefont {I.}~\bibnamefont {{\v{Z}}uti{\'{c}}}},\ }\bibfield  {title} {\bibinfo {title} {Fusion of majorana bound states with mini-gate control in two-dimensional systems},\ }\href {https://doi.org/10.1038/s41467-022-29463-6} {\bibfield  {journal} {\bibinfo  {journal} {Nat. Commun.}\ }\textbf {\bibinfo {volume} {13}} (\bibinfo {year} {2022})}\BibitemShut {NoStop}%
\bibitem [{\citenamefont {Ando}\ \emph {et~al.}(2020)\citenamefont {Ando}, \citenamefont {Miyasaka}, \citenamefont {Li}, \citenamefont {Ishizuka}, \citenamefont {Arakawa}, \citenamefont {Shiota}, \citenamefont {Moriyama}, \citenamefont {Yanase},\ and\ \citenamefont {Ono}}]{Ando2020}%
  \BibitemOpen
  \bibfield  {author} {\bibinfo {author} {\bibfnamefont {F.}~\bibnamefont {Ando}}, \bibinfo {author} {\bibfnamefont {Y.}~\bibnamefont {Miyasaka}}, \bibinfo {author} {\bibfnamefont {T.}~\bibnamefont {Li}}, \bibinfo {author} {\bibfnamefont {J.}~\bibnamefont {Ishizuka}}, \bibinfo {author} {\bibfnamefont {T.}~\bibnamefont {Arakawa}}, \bibinfo {author} {\bibfnamefont {Y.}~\bibnamefont {Shiota}}, \bibinfo {author} {\bibfnamefont {T.}~\bibnamefont {Moriyama}}, \bibinfo {author} {\bibfnamefont {Y.}~\bibnamefont {Yanase}},\ and\ \bibinfo {author} {\bibfnamefont {T.}~\bibnamefont {Ono}},\ }\bibfield  {title} {\bibinfo {title} {Observation of superconducting diode effect},\ }\href {https://doi.org/10.1038/s41586-020-2590-4} {\bibfield  {journal} {\bibinfo  {journal} {Nature}\ }\textbf {\bibinfo {volume} {584}},\ \bibinfo {pages} {373} (\bibinfo {year} {2020})}\BibitemShut {NoStop}%
\bibitem [{\citenamefont {Mayer}\ \emph {et~al.}(2020)\citenamefont {Mayer}, \citenamefont {Dartiailh}, \citenamefont {Yuan}, \citenamefont {Wickramasinghe}, \citenamefont {Rossi},\ and\ \citenamefont {Shabani}}]{Mayer2020}%
  \BibitemOpen
  \bibfield  {author} {\bibinfo {author} {\bibfnamefont {W.}~\bibnamefont {Mayer}}, \bibinfo {author} {\bibfnamefont {M.~C.}\ \bibnamefont {Dartiailh}}, \bibinfo {author} {\bibfnamefont {J.}~\bibnamefont {Yuan}}, \bibinfo {author} {\bibfnamefont {K.~S.}\ \bibnamefont {Wickramasinghe}}, \bibinfo {author} {\bibfnamefont {E.}~\bibnamefont {Rossi}},\ and\ \bibinfo {author} {\bibfnamefont {J.}~\bibnamefont {Shabani}},\ }\bibfield  {title} {\bibinfo {title} {Gate controlled anomalous phase shift in al/{InAs} josephson junctions},\ }\href {https://doi.org/10.1038/s41467-019-14094-1} {\bibfield  {journal} {\bibinfo  {journal} {Nat. Commun.}\ }\textbf {\bibinfo {volume} {11}},\ \bibinfo {pages} {212} (\bibinfo {year} {2020})}\BibitemShut {NoStop}%
\bibitem [{\citenamefont {Dartiailh}\ \emph {et~al.}(2021)\citenamefont {Dartiailh}, \citenamefont {Mayer}, \citenamefont {Yuan}, \citenamefont {Wickramasinghe}, \citenamefont {Matos-Abiague}, \citenamefont {\ifmmode \check{Z}\else \v{Z}\fi{}uti\ifmmode~\acute{c}\else \'{c}\fi{}},\ and\ \citenamefont {Shabani}}]{Dartiailh2021}%
  \BibitemOpen
  \bibfield  {author} {\bibinfo {author} {\bibfnamefont {M.~C.}\ \bibnamefont {Dartiailh}}, \bibinfo {author} {\bibfnamefont {W.}~\bibnamefont {Mayer}}, \bibinfo {author} {\bibfnamefont {J.}~\bibnamefont {Yuan}}, \bibinfo {author} {\bibfnamefont {K.~S.}\ \bibnamefont {Wickramasinghe}}, \bibinfo {author} {\bibfnamefont {A.}~\bibnamefont {Matos-Abiague}}, \bibinfo {author} {\bibfnamefont {I.}~\bibnamefont {\ifmmode \check{Z}\else \v{Z}\fi{}uti\ifmmode~\acute{c}\else \'{c}\fi{}}},\ and\ \bibinfo {author} {\bibfnamefont {J.}~\bibnamefont {Shabani}},\ }\bibfield  {title} {\bibinfo {title} {Phase signature of topological transition in josephson junctions},\ }\href {https://doi.org/10.1103/PhysRevLett.126.036802} {\bibfield  {journal} {\bibinfo  {journal} {Phys. Rev. Lett.}\ }\textbf {\bibinfo {volume} {126}},\ \bibinfo {pages} {036802} (\bibinfo {year} {2021})}\BibitemShut {NoStop}%
\bibitem [{\citenamefont {Baumgartner}\ \emph {et~al.}(2021)\citenamefont {Baumgartner}, \citenamefont {Fuchs}, \citenamefont {Costa}, \citenamefont {Reinhardt}, \citenamefont {Gronin}, \citenamefont {Gardner}, \citenamefont {Lindemann}, \citenamefont {Manfra}, \citenamefont {Junior}, \citenamefont {Kochan}, \citenamefont {Fabian}, \citenamefont {Paradiso},\ and\ \citenamefont {Strunk}}]{Baumgartner2021}%
  \BibitemOpen
  \bibfield  {author} {\bibinfo {author} {\bibfnamefont {C.}~\bibnamefont {Baumgartner}}, \bibinfo {author} {\bibfnamefont {L.}~\bibnamefont {Fuchs}}, \bibinfo {author} {\bibfnamefont {A.}~\bibnamefont {Costa}}, \bibinfo {author} {\bibfnamefont {S.}~\bibnamefont {Reinhardt}}, \bibinfo {author} {\bibfnamefont {S.}~\bibnamefont {Gronin}}, \bibinfo {author} {\bibfnamefont {G.~C.}\ \bibnamefont {Gardner}}, \bibinfo {author} {\bibfnamefont {T.}~\bibnamefont {Lindemann}}, \bibinfo {author} {\bibfnamefont {M.~J.}\ \bibnamefont {Manfra}}, \bibinfo {author} {\bibfnamefont {P.~E.~F.}\ \bibnamefont {Junior}}, \bibinfo {author} {\bibfnamefont {D.}~\bibnamefont {Kochan}}, \bibinfo {author} {\bibfnamefont {J.}~\bibnamefont {Fabian}}, \bibinfo {author} {\bibfnamefont {N.}~\bibnamefont {Paradiso}},\ and\ \bibinfo {author} {\bibfnamefont {C.}~\bibnamefont {Strunk}},\ }\bibfield  {title} {\bibinfo {title} {Supercurrent rectification and magnetochiral effects in symmetric josephson junctions},\ }\href
  {https://doi.org/10.1038/s41565-021-01009-9} {\bibfield  {journal} {\bibinfo  {journal} {Nat. Nanotechnol.}\ }\textbf {\bibinfo {volume} {17}},\ \bibinfo {pages} {39} (\bibinfo {year} {2021})}\BibitemShut {NoStop}%
\bibitem [{\citenamefont {Amundsen}\ \emph {et~al.}()\citenamefont {Amundsen}, \citenamefont {Linder}, \citenamefont {Robinson}, \citenamefont {Žutić},\ and\ \citenamefont {Banerjee}}]{Amundsen2022}%
  \BibitemOpen
  \bibfield  {author} {\bibinfo {author} {\bibfnamefont {M.}~\bibnamefont {Amundsen}}, \bibinfo {author} {\bibfnamefont {J.}~\bibnamefont {Linder}}, \bibinfo {author} {\bibfnamefont {J.~W.~A.}\ \bibnamefont {Robinson}}, \bibinfo {author} {\bibfnamefont {I.}~\bibnamefont {Žutić}},\ and\ \bibinfo {author} {\bibfnamefont {N.}~\bibnamefont {Banerjee}},\ }\href {https://arxiv.org/abs/2210.03549} {\bibinfo {title} {Colloquium: Spin-orbit effects in superconducting hybrid structures}},\ \bibinfo {note} {arXiv.2210.03549, (2022)}\BibitemShut {NoStop}%
\bibitem [{\citenamefont {Tanaka}\ \emph {et~al.}(2022)\citenamefont {Tanaka}, \citenamefont {Lu},\ and\ \citenamefont {Nagaosa}}]{Tanaka2022}%
  \BibitemOpen
  \bibfield  {author} {\bibinfo {author} {\bibfnamefont {Y.}~\bibnamefont {Tanaka}}, \bibinfo {author} {\bibfnamefont {B.}~\bibnamefont {Lu}},\ and\ \bibinfo {author} {\bibfnamefont {N.}~\bibnamefont {Nagaosa}},\ }\bibfield  {title} {\bibinfo {title} {Theory of giant diode effect in $d$-wave superconductor junctions on the surface of a topological insulator},\ }\href {https://doi.org/10.1103/PhysRevB.106.214524} {\bibfield  {journal} {\bibinfo  {journal} {Phys. Rev. B}\ }\textbf {\bibinfo {volume} {106}},\ \bibinfo {pages} {214524} (\bibinfo {year} {2022})}\BibitemShut {NoStop}%
\bibitem [{\citenamefont {Yu}\ \emph {et~al.}(2019)\citenamefont {Yu}, \citenamefont {Ma}, \citenamefont {Cai}, \citenamefont {Zhong}, \citenamefont {Ye}, \citenamefont {Shen}, \citenamefont {Gu}, \citenamefont {Chen},\ and\ \citenamefont {Zhang}}]{Yu2019}%
  \BibitemOpen
  \bibfield  {author} {\bibinfo {author} {\bibfnamefont {Y.}~\bibnamefont {Yu}}, \bibinfo {author} {\bibfnamefont {L.}~\bibnamefont {Ma}}, \bibinfo {author} {\bibfnamefont {P.}~\bibnamefont {Cai}}, \bibinfo {author} {\bibfnamefont {R.}~\bibnamefont {Zhong}}, \bibinfo {author} {\bibfnamefont {C.}~\bibnamefont {Ye}}, \bibinfo {author} {\bibfnamefont {J.}~\bibnamefont {Shen}}, \bibinfo {author} {\bibfnamefont {G.~D.}\ \bibnamefont {Gu}}, \bibinfo {author} {\bibfnamefont {X.~H.}\ \bibnamefont {Chen}},\ and\ \bibinfo {author} {\bibfnamefont {Y.}~\bibnamefont {Zhang}},\ }\bibfield  {title} {\bibinfo {title} {High-temperature superconductivity in monolayer bi2sr2cacu2o8+$\delta$},\ }\href {https://doi.org/10.1038/s41586-019-1718-x} {\bibfield  {journal} {\bibinfo  {journal} {Nature}\ }\textbf {\bibinfo {volume} {575}},\ \bibinfo {pages} {156} (\bibinfo {year} {2019})}\BibitemShut {NoStop}%
\bibitem [{\citenamefont {Wu}\ \emph {et~al.}(2017)\citenamefont {Wu}, \citenamefont {Yuan}, \citenamefont {Meng}, \citenamefont {Chen}, \citenamefont {Sun}, \citenamefont {Chen}, \citenamefont {Dang}, \citenamefont {Tan}, \citenamefont {Liu}, \citenamefont {Yin}, \citenamefont {Zhou}, \citenamefont {Huang}, \citenamefont {Xu}, \citenamefont {Cui}, \citenamefont {Hwang}, \citenamefont {Liu}, \citenamefont {Chen}, \citenamefont {Yan},\ and\ \citenamefont {Peng}}]{Wu2017}%
  \BibitemOpen
  \bibfield  {author} {\bibinfo {author} {\bibfnamefont {J.}~\bibnamefont {Wu}}, \bibinfo {author} {\bibfnamefont {H.}~\bibnamefont {Yuan}}, \bibinfo {author} {\bibfnamefont {M.}~\bibnamefont {Meng}}, \bibinfo {author} {\bibfnamefont {C.}~\bibnamefont {Chen}}, \bibinfo {author} {\bibfnamefont {Y.}~\bibnamefont {Sun}}, \bibinfo {author} {\bibfnamefont {Z.}~\bibnamefont {Chen}}, \bibinfo {author} {\bibfnamefont {W.}~\bibnamefont {Dang}}, \bibinfo {author} {\bibfnamefont {C.}~\bibnamefont {Tan}}, \bibinfo {author} {\bibfnamefont {Y.}~\bibnamefont {Liu}}, \bibinfo {author} {\bibfnamefont {J.}~\bibnamefont {Yin}}, \bibinfo {author} {\bibfnamefont {Y.}~\bibnamefont {Zhou}}, \bibinfo {author} {\bibfnamefont {S.}~\bibnamefont {Huang}}, \bibinfo {author} {\bibfnamefont {H.~Q.}\ \bibnamefont {Xu}}, \bibinfo {author} {\bibfnamefont {Y.}~\bibnamefont {Cui}}, \bibinfo {author} {\bibfnamefont {H.~Y.}\ \bibnamefont {Hwang}}, \bibinfo {author} {\bibfnamefont {Z.}~\bibnamefont {Liu}}, \bibinfo {author} {\bibfnamefont
  {Y.}~\bibnamefont {Chen}}, \bibinfo {author} {\bibfnamefont {B.}~\bibnamefont {Yan}},\ and\ \bibinfo {author} {\bibfnamefont {H.}~\bibnamefont {Peng}},\ }\bibfield  {title} {\bibinfo {title} {High electron mobility and quantum oscillations in non-encapsulated ultrathin semiconducting bi2o2se},\ }\href {https://doi.org/10.1038/nnano.2017.43} {\bibfield  {journal} {\bibinfo  {journal} {Nat. Nanotechnol.}\ }\textbf {\bibinfo {volume} {12}},\ \bibinfo {pages} {530} (\bibinfo {year} {2017})}\BibitemShut {NoStop}%
\bibitem [{\citenamefont {Tewari}\ and\ \citenamefont {Sau}(2012)}]{Sau2012}%
  \BibitemOpen
  \bibfield  {author} {\bibinfo {author} {\bibfnamefont {S.}~\bibnamefont {Tewari}}\ and\ \bibinfo {author} {\bibfnamefont {J.~D.}\ \bibnamefont {Sau}},\ }\bibfield  {title} {\bibinfo {title} {Topological invariants for spin-orbit coupled superconductor nanowires},\ }\href {https://doi.org/10.1103/PhysRevLett.109.150408} {\bibfield  {journal} {\bibinfo  {journal} {Phys. Rev. Lett.}\ }\textbf {\bibinfo {volume} {109}},\ \bibinfo {pages} {150408} (\bibinfo {year} {2012})}\BibitemShut {NoStop}%
\bibitem [{\citenamefont {Sato}\ \emph {et~al.}(2010)\citenamefont {Sato}, \citenamefont {Takahashi},\ and\ \citenamefont {Fujimoto}}]{Sato2010}%
  \BibitemOpen
  \bibfield  {author} {\bibinfo {author} {\bibfnamefont {M.}~\bibnamefont {Sato}}, \bibinfo {author} {\bibfnamefont {Y.}~\bibnamefont {Takahashi}},\ and\ \bibinfo {author} {\bibfnamefont {S.}~\bibnamefont {Fujimoto}},\ }\bibfield  {title} {\bibinfo {title} {Non-abelian topological orders and majorana fermions in spin-singlet superconductors},\ }\href {https://doi.org/10.1103/PhysRevB.82.134521} {\bibfield  {journal} {\bibinfo  {journal} {Phys. Rev. B}\ }\textbf {\bibinfo {volume} {82}},\ \bibinfo {pages} {134521} (\bibinfo {year} {2010})}\BibitemShut {NoStop}%
\bibitem [{\citenamefont {Alidoust}\ \emph {et~al.}(2021)\citenamefont {Alidoust}, \citenamefont {Shen},\ and\ \citenamefont {\ifmmode \check{Z}\else \v{Z}\fi{}uti\ifmmode~\acute{c}\else \'{c}\fi{}}}]{Alidoust2021}%
  \BibitemOpen
  \bibfield  {author} {\bibinfo {author} {\bibfnamefont {M.}~\bibnamefont {Alidoust}}, \bibinfo {author} {\bibfnamefont {C.}~\bibnamefont {Shen}},\ and\ \bibinfo {author} {\bibfnamefont {I.}~\bibnamefont {\ifmmode \check{Z}\else \v{Z}\fi{}uti\ifmmode~\acute{c}\else \'{c}\fi{}}},\ }\bibfield  {title} {\bibinfo {title} {Cubic spin-orbit coupling and anomalous josephson effect in planar junctions},\ }\href {https://doi.org/10.1103/PhysRevB.103.L060503} {\bibfield  {journal} {\bibinfo  {journal} {Phys. Rev. B}\ }\textbf {\bibinfo {volume} {103}},\ \bibinfo {pages} {L060503} (\bibinfo {year} {2021})}\BibitemShut {NoStop}%
\bibitem [{\citenamefont {Luethi}\ \emph {et~al.}(2023)\citenamefont {Luethi}, \citenamefont {Laubscher}, \citenamefont {Bosco}, \citenamefont {Loss},\ and\ \citenamefont {Klinovaja}}]{Luethi2023}%
  \BibitemOpen
  \bibfield  {author} {\bibinfo {author} {\bibfnamefont {M.}~\bibnamefont {Luethi}}, \bibinfo {author} {\bibfnamefont {K.}~\bibnamefont {Laubscher}}, \bibinfo {author} {\bibfnamefont {S.}~\bibnamefont {Bosco}}, \bibinfo {author} {\bibfnamefont {D.}~\bibnamefont {Loss}},\ and\ \bibinfo {author} {\bibfnamefont {J.}~\bibnamefont {Klinovaja}},\ }\bibfield  {title} {\bibinfo {title} {Planar josephson junctions in germanium: Effect of cubic spin-orbit interaction},\ }\href {https://doi.org/10.1103/PhysRevB.107.035435} {\bibfield  {journal} {\bibinfo  {journal} {Phys. Rev. B}\ }\textbf {\bibinfo {volume} {107}},\ \bibinfo {pages} {035435} (\bibinfo {year} {2023})}\BibitemShut {NoStop}%
\bibitem [{\citenamefont {Keimer}\ \emph {et~al.}(2015)\citenamefont {Keimer}, \citenamefont {Kivelson}, \citenamefont {Norman}, \citenamefont {Uchida},\ and\ \citenamefont {Zaanen}}]{Keimer2015}%
  \BibitemOpen
  \bibfield  {author} {\bibinfo {author} {\bibfnamefont {B.}~\bibnamefont {Keimer}}, \bibinfo {author} {\bibfnamefont {S.~A.}\ \bibnamefont {Kivelson}}, \bibinfo {author} {\bibfnamefont {M.~R.}\ \bibnamefont {Norman}}, \bibinfo {author} {\bibfnamefont {S.}~\bibnamefont {Uchida}},\ and\ \bibinfo {author} {\bibfnamefont {J.}~\bibnamefont {Zaanen}},\ }\bibfield  {title} {\bibinfo {title} {From quantum matter to high-temperature superconductivity in copper oxides},\ }\href {https://doi.org/10.1038/nature14165} {\bibfield  {journal} {\bibinfo  {journal} {Nature}\ }\textbf {\bibinfo {volume} {518}},\ \bibinfo {pages} {179} (\bibinfo {year} {2015})}\BibitemShut {NoStop}%
\bibitem [{\citenamefont {Court}\ \emph {et~al.}(2007)\citenamefont {Court}, \citenamefont {Ferguson},\ and\ \citenamefont {Clark}}]{Court_2008}%
  \BibitemOpen
  \bibfield  {author} {\bibinfo {author} {\bibfnamefont {N.~A.}\ \bibnamefont {Court}}, \bibinfo {author} {\bibfnamefont {A.~J.}\ \bibnamefont {Ferguson}},\ and\ \bibinfo {author} {\bibfnamefont {R.~G.}\ \bibnamefont {Clark}},\ }\bibfield  {title} {\bibinfo {title} {Energy gap measurement of nanostructured aluminium thin films for single cooper-pair devices},\ }\href {https://doi.org/10.1088/0953-2048/21/01/015013} {\bibfield  {journal} {\bibinfo  {journal} {Supercond. Sci. Technol.}\ }\textbf {\bibinfo {volume} {21}},\ \bibinfo {pages} {015013} (\bibinfo {year} {2007})}\BibitemShut {NoStop}%
\bibitem [{\citenamefont {Wang}\ \emph {et~al.}(2013)\citenamefont {Wang}, \citenamefont {Ding}, \citenamefont {Fedorov}, \citenamefont {Yao}, \citenamefont {Li}, \citenamefont {Lv}, \citenamefont {Zhao}, \citenamefont {Zhang}, \citenamefont {Xu}, \citenamefont {Schneeloch}, \citenamefont {Zhong}, \citenamefont {Ji}, \citenamefont {Wang}, \citenamefont {He}, \citenamefont {Ma}, \citenamefont {Gu}, \citenamefont {Yao}, \citenamefont {Xue}, \citenamefont {Chen},\ and\ \citenamefont {Zhou}}]{Wang2013}%
  \BibitemOpen
  \bibfield  {author} {\bibinfo {author} {\bibfnamefont {E.}~\bibnamefont {Wang}}, \bibinfo {author} {\bibfnamefont {H.}~\bibnamefont {Ding}}, \bibinfo {author} {\bibfnamefont {A.~V.}\ \bibnamefont {Fedorov}}, \bibinfo {author} {\bibfnamefont {W.}~\bibnamefont {Yao}}, \bibinfo {author} {\bibfnamefont {Z.}~\bibnamefont {Li}}, \bibinfo {author} {\bibfnamefont {Y.-F.}\ \bibnamefont {Lv}}, \bibinfo {author} {\bibfnamefont {K.}~\bibnamefont {Zhao}}, \bibinfo {author} {\bibfnamefont {L.-G.}\ \bibnamefont {Zhang}}, \bibinfo {author} {\bibfnamefont {Z.}~\bibnamefont {Xu}}, \bibinfo {author} {\bibfnamefont {J.}~\bibnamefont {Schneeloch}}, \bibinfo {author} {\bibfnamefont {R.}~\bibnamefont {Zhong}}, \bibinfo {author} {\bibfnamefont {S.-H.}\ \bibnamefont {Ji}}, \bibinfo {author} {\bibfnamefont {L.}~\bibnamefont {Wang}}, \bibinfo {author} {\bibfnamefont {K.}~\bibnamefont {He}}, \bibinfo {author} {\bibfnamefont {X.}~\bibnamefont {Ma}}, \bibinfo {author} {\bibfnamefont {G.}~\bibnamefont {Gu}}, \bibinfo {author}
  {\bibfnamefont {H.}~\bibnamefont {Yao}}, \bibinfo {author} {\bibfnamefont {Q.-K.}\ \bibnamefont {Xue}}, \bibinfo {author} {\bibfnamefont {X.}~\bibnamefont {Chen}},\ and\ \bibinfo {author} {\bibfnamefont {S.}~\bibnamefont {Zhou}},\ }\bibfield  {title} {\bibinfo {title} {Fully gapped topological surface states in bi2se3 films induced by a d-wave high-temperature superconductor},\ }\href {https://doi.org/10.1038/nphys2744} {\bibfield  {journal} {\bibinfo  {journal} {Nat. Phys.}\ }\textbf {\bibinfo {volume} {9}},\ \bibinfo {pages} {621} (\bibinfo {year} {2013})}\BibitemShut {NoStop}%
\end{thebibliography}%
\end{document}